\documentclass[%
 reprint,
 floatfix,
preprintnumbers,
 amsmath,amssymb,
 aps,nofootinbib,prd,
]{revtex4-2}

\usepackage{scrextend}
\usepackage{graphicx}
\usepackage{dcolumn}
\usepackage{bm}
\usepackage{xcolor}
\usepackage{siunitx}
\usepackage{comment}
\usepackage{mwe}
\usepackage{hyperref}
\hypersetup{
  colorlinks = true,
  urlcolor=teal,
  citecolor=blue,
  linkcolor=red
}

\newcommand{\mudhi}{%
  \begingroup\normalfont
  \raisebox{-4pt}{
  \includegraphics[height=13pt]{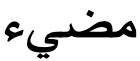}%
  }
  \endgroup
}

\newcommand{\mudhiIPA}{%
  \begingroup\normalfont
  \raisebox{-2.5pt}{
  \includegraphics[height=9pt]{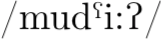}%
  }
  \endgroup
}

\newcommand{\non}{\ensuremath{N_{\rm on}} \xspace}
\newcommand{\noff}{\ensuremath{N_{\rm off}} \xspace}

\newcommand{\lik}{\ensuremath{\mathcal{L}} \xspace}
\newcommand{\hbc}{\ensuremath{\hat{B}_c}\xspace}

\providecommand*{\eu}%
{\ensuremath{\mathrm{e}}}

\usepackage{bm}
\usepackage{bbold}
\usepackage{amsbsy}
\usepackage{subfigure}
\usepackage{slashed}
\usepackage{afterpage}
\usepackage{psfrag}
\usepackage{dsfont}
\usepackage{soul}
\usepackage{appendix}
\usepackage{lineno}
\begin{document}


\title{Search for dark photons using a multilayer dielectric haloscope\\ equipped with a single-photon avalanche diode}

\author{Laura Manenti} \email{laura.manenti@nyu.edu}
\author{Umang Mishra}
\author{Gianmarco Bruno}
\author{Henry~Roberts}
\author{Panos Oikonomou}
\author{Renu Pasricha}
\author{Isaac Sarnoff}
\author{James Weston}
\author{Francesco Arneodo}
\affiliation{Division of Science, New York University Abu Dhabi, United Arab Emirates}
\affiliation{Center for Astro, Particle and Planetary Physics (CAP$^3$), New York University Abu Dhabi, United Arab Emirates}

\author{Adriano Di Giovanni}
\affiliation{Gran Sasso Science Institute (GSSI), Via Iacobucci 2, I-67100 L’Aquila, Italy}
\affiliation{Istituto Nazionale di Fisica Nucleare (INFN) - Laboratori Nazionali del Gran Sasso, I-67100 Assergi, L’Aquila, Italy}
\affiliation{Center for Astro, Particle and Planetary Physics (CAP$^3$), New York University Abu Dhabi, United Arab Emirates}

\author{Alexander John Millar}
\affiliation{The Oskar Klein Centre, Department of Physics, Stockholm University, AlbaNova, SE-10691 Stockholm, Sweden}
\affiliation{Nordita, KTH Royal Institute of Technology and
Stockholm University, Roslagstullsbacken 23, 10691 Stockholm, Sweden}

\author{Knut Dundas Morå}
\affiliation{%
 Physics Department, Columbia University, New York, New York 10027, USA}%

\date{\today}

\begin{abstract}
We report on the results of the search for dark photons with mass around \SI{1.5}{eV/c^2} using a multilayer dielectric haloscope equipped with an affordable and commercially available photosensor. The multilayer stack, which enables the conversion of dark photons (DP) to Standard Model photons, is made of 23 bilayers of alternating SiO$_2$ and Si$_3$N$_4$ thin films with linearly increasing thicknesses through the stack (a configuration known as a ``chirped stack''). The thicknesses have been chosen according to an optimisation algorithm in order to maximise the DP-photon conversion in the energy region where the photosensor sensitivity peaks.
This prototype experiment, baptised MuDHI \mudhi \;(Multilayer Dielectric Haloscope Investigation) by the authors of this paper, has been designed, developed and run at the Astroparticle Laboratory of New York University Abu Dhabi, which marks the first time a dark matter experiment has been operated in the Middle East.

No significant signal excess is observed, and the method of maximum log-likelihood is used to set exclusion limits at 90\% confidence level on the kinetic mixing coupling constant between dark photons and ordinary photons.
\end{abstract}

\maketitle
  \preprint{NORDITA-2021-087}   

\section{\label{sec:level1} Introduction}
There is compelling evidence that about 85\% of the matter content in the Universe is composed of dark matter (DM), a new type of beyond-the-Standard-Model particle that has not yet been detected~\cite{aghanim2020planck}. While candidates like weakly-interacting massive particles (WIMPs) have been the prime target for dark matter experiments over the past decades~\cite{Brink_2009, Aprile_2010}, new particles like axions and dark photons are increasingly gaining traction. 

In this work we search for dark matter with mass around \SI{1.5}{eV/c^2} in the form of dark photons (DPs), theorised particles which belong to the so-called dark sector (termed dark because they are neutral under the Standard Model interactions). Although the existence of dark matter is motivated  by its gravitational coupling to the Standard Model, most of the experiments searching for DM do so based on its presumed weak interaction. Similarly, it is assumed that the detection of DPs is experimentally achievable because of the interaction between the SM and the dark sector through a vector portal. In the vector portal, the interaction is possible because of the kinetic mixing between one dark boson, the dark photon arising from an extra U(1) symmetry, and one ordinary boson, which is taken to be the SM photon of the U(1) gauge group of electromagnetism~\cite{Fabbrichesi_2021}. The kinetic-mixing leads to DP-photon oscillations, very much like what happens between neutrino masses and flavour eigenstates. The Lagrangian for the dark photon with vector field $\tilde X_{\mu}$ and dark photon tensor $\tilde F'_{\mu\nu}$ in the propagation basis (i.e.~the mass basis) is~\cite{Arias_2012, Fabbrichesi_2021, An_2015_xe10, holdom1986two}
\begin{equation}
\label{eq:lagrangian}
    \begin{aligned}
    \mathcal{L} = -\frac{1}{4} \tilde F_{\mu\nu} \tilde F^{\mu\nu} 
        - \frac{1}{4}\tilde F'_{\mu\nu}\tilde F'^{\mu\nu} &
        + \frac{\chi}{2}\tilde F_{\mu\nu}\tilde F'^{\mu\nu} \\
        + \frac{{m_X}^2}{2}\tilde X_{\mu}\tilde X^{\mu}
        + e J^\mu\tilde A_\mu \, ,
        \end{aligned}
\end{equation}
where $\tilde F_{\mu\nu}$ is the electromagnetic field tensor, $\chi$ is the mixing coupling constant, $m_X$ is the mass of the dark photon, $e$ is the electric charge, $J_{\mu}$ is the current density vector, and $\tilde A_{\mu}$ is the electromagnetic four-potential. Note that in Eq.~\ref{eq:lagrangian} we have only considered leading-order terms in $\chi$. For the full expression see, for example, Ref.~\cite{caputo2021dark}. 

The technique we exploit for the detection of dark photons makes use of a stack of alternating dielectric layers and a receiver. This type of detector is dubbed ``dielectric haloscope'' (or simply ``haloscope'') and has been employed to search for axions in the mass range of 40--\SI{400}{\micro eV/c^2}~\cite{Caldwell_2017, Li:2020ogf} (when coupled to an external magnetic field) and proposed for the detection of dark photons in the 0.1--\SI{10}{eV/c^2} mass range~\cite{Baryakhtar_2018}. 

We do not observe an excess in the data, and we place constraints on the kinetic mixing coupling constant between dark photons and ordinary photons at 90\% confidence level (CL) in the mass region under investigation. We do so using a multilayer dielectric haloscope equipped with an affordable and commercially available photosensor.
The materials used for the stack of 23 bilayers are SiO$_2$ and Si$_3$N$_4$. We explore the challenges of making, characterising, and operating a dielectric stack with a high number of layers. 
An aspherical lens is used to concentrate the emitted power onto a single-photon avalanche diode (SPAD). 

This prototype detector has been designed, developed and run at New York University Abu Dhabi, which also marks the first time a dark matter experiment has been operated in the Middle East. This proof-of-principle experiment has been named MuDHI (Multilayer Dielectric Haloscope Investigation)--in Arabic, the word mudhì\footnote{Pronounced\,\mudhiIPA in IPA.} means ``luminous.''
Although peer-reviewed results have yet to be reported, the authors became aware of an existing haloscope prototype which used five bilayers and a superconducting nanowire single-photon detector as a photosensor in the final stages of preparing this manuscript. These findings are available at Ref.~\cite{chiles2021constraints}.

The paper is organised as follows: in Sec.~\ref{sec:sec2} we give an overview of the experiment design; in Sec.~\ref{sec:sec3} we explain the theory behind the conversion of dark photons into ordinary photons; in Sec.~\ref{sec:sec4} we describe in detail our specific experimental setup; in Sec.~\ref{sec:sec5} we report on the results of the experiment; and finally, we summarise and conclude in Sec.~\ref{sec:sec6}. All the plots created for this article can be reproduced using the code available at \url{https://github.com/arneodoslab/haloscope.git}. 

\section{\label{sec:sec2} Experiment design}
\begin{figure}[tb]
	\begin{center}
	\includegraphics[width=0.4
	\textwidth]{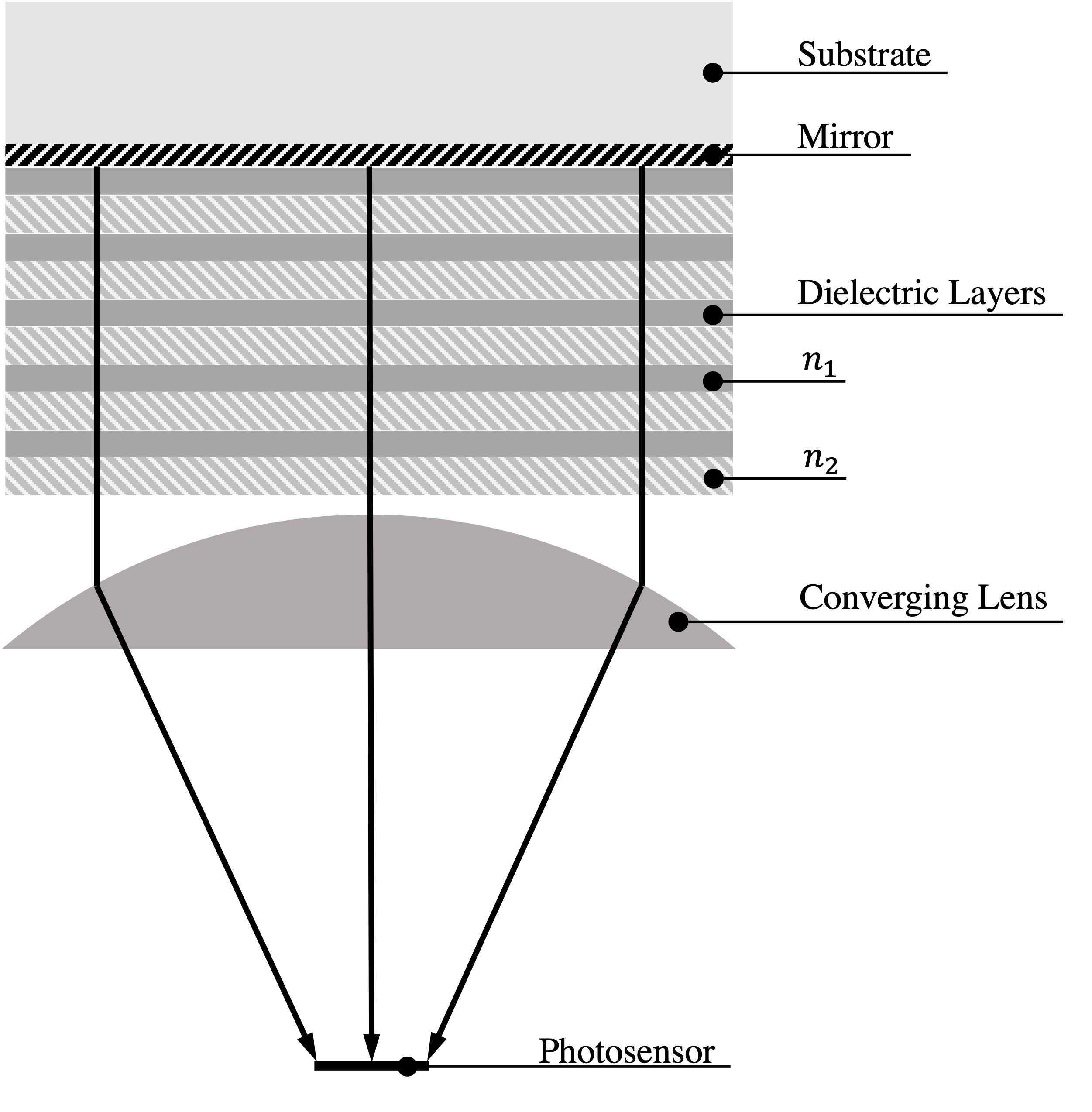}
	\caption{Schematic of a multilayer dielectric haloscope. The bilayers are characterised by the indices of refraction $n_1$ and $n_2$. The mirror layer is optional.}
	\label{fig:schematic_haloscope}
	\end{center}
\end{figure}
The haloscope, shown schematically in Fig.~\ref{fig:schematic_haloscope}, is composed of three main parts: the stack, which enables the conversion from dark photons to photons, the lens, which focuses the converted photons, and the photosensor.                                             
The design of the haloscope is a reverse-engineering process as the first component to be chosen is not the stack but the photosensor, which will determine the frequency of optimal detection efficiency. This has to be capable of single-photon detection, needs to have a small active area (comparable to the size of the spot at the focal point of the lens), very low dark count rate and high detection efficiency. It also must guarantee stable operation over several hours.

Given the aforementioned requirements, our choice fell on single-photon avalanche diodes, SPADs (more in Sec.~\ref{sec:sec4b}). We searched for a SPAD whose photodetection efficiency was peaked around \SI{800}{nm}, i.e.~about \SI{1.5}{eV}, as it corresponds to an under-explored DP mass range of parameter space (note that in the DP-photon kinetic mixing, the rest mass of the DP is entirely converted into the energy of the photon). Initially, we employed a modified \SI{0.1}{mm} diameter version of the Hamamatsu S12426-02 Si APD~\cite{Hamamatsu_datasheet}. Although Hamamatsu did not have experience with the evaluation of this sensor in Geiger mode, the company was willing to provide two samples especially fabricated for our needs. The sensors had a quantum efficiency (i.e.~the probability of an incident photon creating an electron-hole pair in the sensitive volume of the sensor) peaked at \SI{809}{nm}. 
In parallel, we checked for the availability of two dielectric materials with low and high refractive indices, respectively, that would be transparent at the wavelength of interest, and a commercially available technique that would allow their deposition (see Sec.~\ref{sec:sec3a} for more details). We picked SiO$_2$ and Si$_3$N$_4$, that can be deposited in thin layers using a process called plasma enhanced chemical vapour deposition (PECVD)~\cite{feridun_04, shi_19, li_2019}. Finally, the thicknesses of the stack bilayers were chosen so that the conversion from dark photon to ordinary photon as a function of mass would be peaked where the QE of the Hamamatsu SPAD was maximised.
Unfortunately, during the photosensor characterisation, it became clear that the photosensor did not work properly  when operated in Geiger mode. As such, although the stack had been already manufactured according to the Hamamatsu QE curve, we were forced to switch to another photosensor. We first tested the haloscope with a SPAD from Laser Components (model SAP500), but despite its low dark count rate, it could not be operated stably when cooled down (the Peltier cell that we added to the setup to cool down the SPAD did not allow for a constant operating temperature). Eventually, the photosensor used for our final test was the C30902SH-TC from Excelitas with a single stage (-TC) thermoelectric cooler (TEC) incorporated for better noise performance at lower temperatures, an active area of \SI{0.5}{mm} diameter and PDE maximum at $\approx$\SI{800}{nm}.

To find the appropriate lens, which needs to be transparent to the wavelength chosen, we ran a GEANT4 simulation (see appendix~\ref{appendix:GEANT4}) and subsequently checked the results against the commercially available options. 
In the next section, we outline the physics behind the working principle of the dielectric haloscope. 

\section{\label{sec:sec3} Dark photon conversion in dielectric haloscopes} 
To explore the mixing of dark photons and visible photons, we can rewrite the Lagrangian in Eq.~\eqref{eq:lagrangian} in the interaction basis in the limit of small $\chi$
\begin{align}\label{eq:Lagrangian2}
\mathcal{L} \supset & -\dfrac{1}{4} {{F}}_{\mu \nu} {F}^{\mu \nu} - \dfrac{1}{4}{F'}_{\mu \nu} {F'}^{ \mu \nu} + e  J^{\mu}{A}_{\mu}
\notag \\
& + \dfrac{m_{X}^2}{2}\left({X}^{\mu} {X}_{\mu} +2\chi{X}_\mu {A}^\mu\right) \,.
\end{align}
In writing this, we have used that for very small $\chi$ we can simply make a replacement $\tilde X^\mu\to X^\mu-\chi A^\mu$ while $\tilde A^\mu\to A^\mu$. For a more detailed treatment, see for example Ref.~\cite{Gelmini:2020kcu}. Since the propagation and interaction basis are distinct, a propagating dark photon state is a mixture of the dark photon and the SM photon interaction states. This implies that a dark photon possesses a small electric field $E_X$. 

Given the current DM relic abundance, for high DP masses ($\mathcal{O}({\rm eV})$) the occupation number may not be high, so one should adopt a quantised approach to study the oscillation between DPs and ordinary photons. However, according to Refs.~\cite{Raffelt:1991ck,Ioannisian:2017srr}, a classical field procedure (where by ``classical field'' we mean the expectation value of the field after evolution and upon measurement) is still a viable method if one is only interested in the average DP-SM photon conversion rate and not in higher order quantum corrections. Here, we shall opt for the classical field procedure.

In principle the DP field can be complicated, with non-trivial structure, velocities and direction. Such complexities can be captured via a Fourier transform (e.g. Refs.~\cite{Knirck:2018knd,Gelmini:2020kcu}), but
for the sake of simplicity, we shall consider the DP at a given point in time in the detector to be  described by a plane wave with zero momentum, and solve for the propagating eigenstates. As we are concerned with dielectric media, we can rewrite the bound charges contained in $J^{\mu}$ into the refractive index $n$ of the material, and following the prescription in Ref.~\cite{Jaeckel:2013eha} we can write down the equations of motion in media of the DP and photon fields
\begin{widetext}
\begin{align}
    \left [(\omega^2-k^2)\left (\begin{array}{cc} 1 & 0 \\0 & 1  \end{array}\right )-\omega^2(1-n^2)\left(\begin{array}{cc} 1 & 0 \\0 & 0  \end{array}\right)-m^2_X\left (\begin{array}{cc} 0 & -\chi \\-\chi & 1  \end{array}\right)\right]\left (\begin{array}{cc} \vec{A}  \\\vec{X}  \end{array}\right )=0 \,,
\end{align}
\end{widetext}
where we have neglected terms of $\mathcal{O}(\chi^2)$ and used the fact that the DP velocity is small ($v\simeq 10^{-3}$)\footnote{Natural units are assumed throughout this work unless otherwise stated.} to choose a gauge with $A^0\simeq X^0\simeq 0$~\cite{Jaeckel:2013sqa}.
Solving for the propagating eigenstates, which correspond to a propagating DP-like and a photon-like state, we obtain
\begin{subequations}
\begin{align}
   \hbox to 5em{DP-like: \hfil}~\left (\begin{array}{cc}\vec{A}  \\\vec{X}  \end{array}\right )&= \vec{X_0}\left (\begin{array}{cc} -\chi_{\rm eff} \\1 \end{array}\right)e^{i(\vec{k}\cdot\vec{x}-\omega t)}\\
   \hbox to 5em{photon-like: \hfil}~\left (\begin{array}{cc} \vec{A}  \\\vec{X} \end{array}\right )&= \vec{A_0}\left(\begin{array}{cc}  1\\\chi_{\rm eff} \end{array}\right)e^{i(\vec{k}\cdot\vec{x}-\omega t)}\,,
\end{align}
\end{subequations}
where
\begin{equation}
    \chi_{\rm eff}=\chi\frac{m_X^2}{m_X^2-\omega^2+n^2\omega^2}\simeq\frac{\chi}{n^2}\,,
\end{equation}
is the effective kinetic mixing angle in media. 
The propagating DP-like state has corresponding energy density (i.e.~local dark matter density) given by
\begin{equation}
    \rho \simeq \frac{m_X^2}{2}|\vec{X}_0|^2 \, .
\end{equation}
We can now write down the $E$-field induced by the DP-like state\footnote{Only the photon interaction component of the propagating DP-like state contributes to the electromagnetic field.} in an infinite medium, $\vec{E}_X$, which is given by
\begin{equation}
   \vec{E}_X = -i\frac{\chi m_X}{n^2}\vec{X}_0 \, .
\end{equation}
One can compare this equation with the case more often studied in dielectric haloscopes, the axion $a$, which, for a mass $m_a$ with coupling $g_{a\gamma}$ in an external magnetic field $\vec{B}_{\rm e}$, is given by
\begin{equation}
\vec{E}_a=-\frac{g_{a\gamma}\vec{B}_{\rm e}}{n^2}a\,.
\end{equation}
Unlike the axion case, DP induced $E$-fields can point in any direction, and the polarisation may actually have a non-trivial cosmological structure~\cite{Arias:2012az, caputo2021dark}. 

\subsection{\label{sec:sec3a} Multilayer dielectric stack}
\begin{figure}[tb]
	\begin{center}
	\includegraphics[width=0.4
	\textwidth]{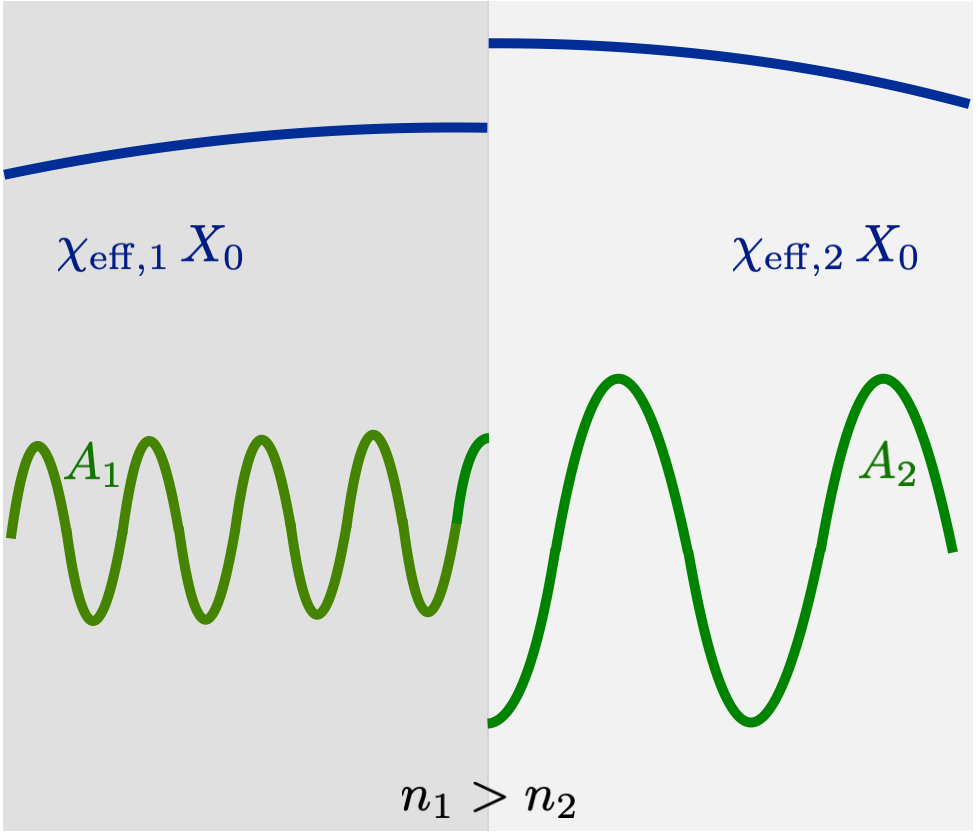}
	\caption{Schematic of a DP like state of amplitude $X_0$ interacting with an interface between two mediums with refractive indices $n_1$ and $n_2$. The DP induced $E$-fields are proportional to the effective mixing angle $\chi_{\rm eff}$, which depends on the refractive index and so changes across the interface. Due to the discontinuity in the DP induced $E$-field (dark blue lines), photon-like propagating waves (green) labeled by $A_{1,2}$ are emitted to compensate. The DP wavelength is exaggerated so as to be visible. }
	\label{fig:interface}
	\end{center}
\end{figure}
While a non-relativistic propagating DP-like state can generate a small field $E_X$ in an infinite medium, it cannot convert to an ultra-relativistic propagating photon-like state due to conservation of momentum. One interesting exception to this is when $m_X$ matches the plasma frequency of a material, as in plasma haloscopes~\cite{Lawson:2019brd,Gelmini:2020kcu}. Alternatively, one can mix the photon with a quasi-particle axion to provide a massive photon-like state~\cite{Marsh:2018dlj,Schutte-Engel:2021bqm}. However, if one breaks translation invariance, for example by having a change in the medium, one can generate real propagating photon-like states~\cite{Horns:2012jf}. To demonstrate this effect, we follow the calculation in Ref.~\cite{Jaeckel:2013eha}.

Consider a system with two half infinite isotropic and homogeneous media, with refractive indices $n_1$ and $n_2$ and the interface occurring at $x=0$, as shown in Fig.~\ref{fig:interface}. If only the propagating DP-like state is present, one would have a discontinuous electric field $E_X$ across the boundary as $\chi_{\rm eff}$ changes. However, Maxwell's equations enforce the boundary conditions
\begin{subequations}
\begin{align}
{\bf E}_{||,1}&={\bf E}_{||,2}\,,\\
{\bf H}_{||,1}&={\bf H}_{||,2}\,.
\end{align}\label{eq:boundary}
\end{subequations}
In other words, the parallel $E$- and $H$-field components must be conserved at the interface. Because of this, propagating photon-like states must also be generated in order to satisfy the boundary conditions. We can solve for the $E$-fields by allowing for a left moving wave in medium~1 and a right moving wave in medium~2, with amplitudes $A_1$ and $A_2$, respectively. We can write the fields in medium 1 as
\begin{equation}
    \vec{X}_0\left(\begin{array}{cc} -\chi_{\rm eff,1} \\1 \end{array}\right)e^{-i\omega t}+\vec A_1\left(\begin{array}{cc}  1\\ \chi_{\rm eff,1}  \end{array}\right)e^{-i(n_1\omega x+\omega t)}\,
\end{equation}
and the fields in medium 2 as
\begin{equation}
    \vec{X}_0\left(\begin{array}{cc} -\chi_{\rm eff,2}  \\1 \end{array}\right)e^{-i\omega t}+\vec A_2\left(\begin{array}{cc}  1\\ \chi_{\rm eff,2} \end{array}\right)e^{i(n_2\omega x-\omega t)}\,.
\end{equation}
For simplicity we neglect the DP velocity, however this being non-zero will make generated photon waves not entirely perpendicular to the interface, typically giving an opening angle of ${\cal O}(10^{-3}\, \rm rad)$~\cite{Jaeckel:2013sqa}.  
Using the boundary conditions in Eq.~\eqref{eq:boundary} we can then solve for $\vec{A}_1$ and $\vec{A}_2$ to obtain~\cite{Jaeckel:2013eha}:
\begin{subequations}
\begin{align}
    \vec{A}_1&=\chi\left (\frac{1}{n_2^2}-\frac{1}{n_1^2} \right )\frac{n_2}{n_1+n_2}\vec X_0^{||}\,,\\
      \vec{A}_2&=-\chi\left (\frac{1}{n_2^2}-\frac{1}{n_1^2} \right )\frac{n_2}{n_1+n_2}\vec X_0^{||}\,,
\end{align}\label{eq:dish}
\end{subequations}
where $\vec X_0^{||}\equiv\vec X_0-\left (\hat x\cdot \vec X_0\right )\hat x$. We see that for a DP with a polarisation perpendicular to the interface no conversion takes place. 

As the literature has tended to focus on axions rather than DP, it is instructive to compare the above expressions with the $E$-fields generated by an axion when $\vec{ B}_e$ is parallel to the interface~\cite{Millar:2016cjp}
\begin{subequations}
\begin{align}
    i\vec{A}^{\rm axion}_1&=\frac{g_{a\gamma}\vec{B}_{\rm e}a}{\omega}\left (\frac{1}{n_2^2}-\frac{1}{n_1^2} \right )\frac{n_2}{n_1+n_2}\,,\\
     i \vec{A}^{\rm axion}_2&=-\frac{g_{a\gamma}\vec{B}_{\rm e}a}{\omega}\left (\frac{1}{n_2^2}-\frac{1}{n_1^2} \right )\frac{n_2}{n_1+n_2}\,.
\end{align}
\end{subequations}
The structure is exactly the same, except for the unknown DP polarisation. To convert the amplitudes of the two calculations one can make this simple substitution
\begin{equation}
\label{eq:chitoaxion}
    \chi=\frac{g_{a\gamma}B_{\rm e}}{\omega}\cos\theta\,,
\end{equation}
where $|\cos\theta|=X_0^{||}/X_0$ and $\theta$ is the angle between the polarisation $\hat X_0$ and the plane of the interface (i.e the complementary angle between the polarisation and the vector normal to the surface).
In the simple limit of a transition from a mirror to vacuum, one recreates a dish antenna experiment~\cite{Horns:2012jf}. For example if we take $n_1\to\infty$ and $n_2=1$ the flux of the propagating photon-like state, $\Phi_{\rm DA}$, is given by
\begin{equation}
    \Phi_{\rm DA}=\frac{E_2^2}{2\omega}=\frac{\rho}{\omega}\chi^2\cos^2\theta\,,
\end{equation}
where $E_2=\omega A_2$. 
There also remains the energy stored in the $E$-field component of the DP-like propagation state~\cite{Millar:2016cjp}. However, as DP-like propagation states do not bend in medium like photon-like states, they are not focused by a lens, and so should be negligible at the photodetector. 

One downside of the dish antenna approach is that while the device works over a very wide range of frequencies, it is not optimised for any particular frequency. As pointed out in Ref.~\cite{Jaeckel:2013eha,Millar:2016cjp}, one can actually enhance the conversion of DP to photons by having many layers arranged to achieve constructive interference: a multilayer dielectric haloscope. While originally proposed to have movable dielectric disks in order to facilitate a scan over frequency space, at high frequencies it is more practical to have a set of thin films~\cite{Baryakhtar_2018}. While it is not possible to scan by changing layer thicknesses, either large stacks of many layers can be arranged to cover wider frequency spaces, or the stacks can be swapped out to search a different parameter space.  

Such a system can be described either classically using transfer matrices~\cite{Millar:2016cjp}, or using an overlap integral easily derived through a quantum field theory approach~\cite{Ioannisian:2017srr}. As long as we are only concerned with the expectation value of the rate of conversion, both calculations will give the same final answer~\cite{Ioannisian:2017srr}. In either case, the enhancement due to constructive interference is usually referred to as a ``boost factor'', defined so that the outgoing flux of a dielectric haloscope is related to the flux of a dish antenna by
\begin{equation}
    \label{eq:dhflux}
    \Phi_{\rm DH}=\beta^2(\omega)\Phi_{\rm DA}\,.
\end{equation}
To calculate the boost factor we use the one dimensional transfer matrix formalism developed in Ref.~\cite{Millar:2016cjp}.

In principle the finite size of the device along with the non-zero velocity of the dark photon can lead to deviations from the one dimensional formalism. This can occur in two ways. Dark photon velocities transverse to the interfaces lead to photons being emitted from the stack at a small angle (i.e.~not perpendicular to the surface), as well as a frequency shift on the order of $\delta\omega/\omega\simeq 10^{-6}$~\cite{Millar:2017eoc}. 
Similarly, the finite size of the layers will also break translation invariance on the stack plane and provide some transverse momentum to the disks. However, as the layers are much wider than a de Broglie wavelength, as long as the layers are smooth (i.e.~flat across the surface at a similar level to the error in the layer thicknesses) this effect will be subdominant. The second possible effect of a finite DP velocity is a change of phase as the DP passes through the different layers of the stack. However, such effects only become significant when the length of the stack is ${\cal O}(10\%)$ of the DP de Broglie wavelength~\cite{Millar:2017eoc}. For this effect to influence our setup it would need to consist of hundreds of layers.

To properly include effects of finite size and surface imperfections, one should do a 3D analysis in the vein of Ref.~\cite{Knirck:2019eug,MADMAX:2021lxf}. As the priority  of this work is to explore a prototype design of a dielectric haloscope, we will use the one dimensional formalism and simply note that the requirements on surface roughness are of a similar order as the requirement for layer thickness to be known precisely~\cite{MADMAX:2021lxf}.

\subsection{\label{sec:sec3b} Expected rate and boost factor}
To estimate the rate expected for a given dielectric haloscope, we can rewrite Eq.~\eqref{eq:dhflux} more explicitly to see
\begin{equation}
\begin{split}
\label{eq:expectedrate}
    &R_{\rm DH} = \Phi_{\rm DH} A = \\
    &= \frac{1.17}{\rm day}\frac{A}{\rm cm^2}\left ( \frac{\chi}{10^{-12}}\right )^2\frac{\rho}{0.45\,{\rm GeV}/{\rm cm}^3}\frac{\rm eV}{m_X}\beta^2\cos^2\theta\,,
\end{split}
\end{equation}
where $A$ is the area of the stack. To increase the signal, one must either increase $A$ or increase the boost factor. As this flux must be detected over some background counts, one can also increase the signal to noise ratio by having a lower dark count photosensor, which plays a large role in both choosing photodetector technology and the operating frequency. One can write the signal to noise ratio over a period of time $T$ as as~\cite{doi:10.1142/S0217732398003442,Bityukov:2000tt}
\begin{equation}
    \frac{\rm S}{\rm N}=2(\sqrt{N_{\rm S}+N_{\rm B}}-\sqrt{N_{\rm B}})\,,
\end{equation}
where $N_{\rm S}$ and $N_{\rm B}$ are the signal and background counts over the time $T$. On average the counts can be estimated from the signal count rate $n_{\rm S}$ and the dark count rate $n_{\rm B}$ via $N_{\rm S} = n_{\rm S}T$ and $N_{\rm B} = n_{\rm B}T$. Thus, we can see that the significance of a measurement increases as $\sqrt{T}$, and therefore that one gains from having a signal measurement (also referred to as ``on measurement'' in this report) extended over a long time.

Perhaps the least trivial parameter to tune is the boost factor. While increasing measurement time and area results in a direct increase in the expected signal, constructive interference can only be achieved over a finite range of frequencies. In particular, it was noted in Ref.~\cite{Millar:2016cjp} that 
\begin{equation}
    \int d\omega\,\beta^2\propto N\,,
\end{equation} 
where $N$ is the number of layers. In other words, for a fixed number of layers, one can only gain in $\beta$ by sacrificing the frequency range width over which the DP to photon conversion would be enabled. Moreover, large boost factors generally have a correspondingly high sensitivity to error, as they require either more resonant behaviour, or many layers all aligned for constructive interference~\cite{Millar:2016cjp}. In any case, while in principle $N$ may be increased at the expense of a smaller frequency range width covered, in practice this is not a trivial task. Firstly, for large thicknesses ($\approx$\SI{10}{\micro\meter} and above), a high compressing
stress and the fact that the deposition has to be paused twice or more times to allow for the cleaning of the PECVD reactor may cause cracking of the deposited layers. Secondly, with increasing number of layers the measurement of the stack becomes more difficult. Another consideration that one should take into account is that the presence of more than a few hundred layers may lead to DP dephasing over the thickness of the device~\cite{Millar:2017eoc}.

The expected rate also depends on the polarisation structure of the DP, as shown by the presence of the $\cos^2\theta$ term in Eq.~\eqref{eq:expectedrate}. Depending on the cosmological evolution of the DP, it may be possible for the polarisation to be essentially random in each coherence patch, have structures on large scales, or even potentially be almost constant everywhere. In this paper, we shall consider the simple case of a totally randomised polarisation, which, after averaging over many coherence times, gives
\begin{equation}
    \label{eq:costhetaavg}
    \langle \cos^2\theta\rangle=\frac{2}{3}\,.
\end{equation}
For the derivation of Eq.~\eqref{eq:costhetaavg} see appendix~\ref{appendix:cos-derivation}. For the opposite extreme, i.e.~the DP is polarised along a definite direction, one should ideally do a full analysis for which one needs to use directional and timing information~\cite{caputo2021dark}. As the Earth rotates, $\langle \cos^2\theta\rangle$ will change, meaning that the resulting signal will be time varying and directional. For a conservative estimate, one can treat the measurement as instantaneous and choose an angle $\theta$ such that at least $95\%$ of angles are larger. Such a procedure would give $\langle\cos^2\theta\rangle=0.0975$, however we must stress that with detailed timing information and an optimised measurement strategy, the sensitivity in the constant polarisation scenario can be made comparable to the randomised polarisation case~\cite{caputo2021dark}.

\section{\label{sec:sec4}Experimental setup}

\subsection{\label{sec:sec4a1} A 23-bilayer SiO2/Si3N4 mirrored chirped stack}
As mentioned in Section~\ref{sec:sec2}, the dielectric stack is composed of alternating layers of SiO$_2$ and Si$_3$N$_4$ where the choice of the materials is dictated by theoretical and experimental constraints. On the theoretical front, Eq.~\eqref{eq:dish} shows that a high refractive index contrast between the two materials enhances the DP-SM photon conversion. Silicon dioxide and silicon nitride have refractive indices of 1.45 and 2.02 at \SI{800}{nm} respectively, which makes them suitable candidate materials. From an experimental point of view, it is important that the chosen materials are transparent in the region of interest, i.e.~the frequency range over which the photosensor is sensitive. The Hamamatsu photosensor initially chosen for the experiment had a QE peaked around \SI{809}{nm}. Both SiO$_2$ and Si$_3$N$_4$ are transparent in this wavelength region, making them good candidates in this regard as well~\cite{wang_17, karouta_2012}.

Another fundamental consideration is whether there exists a technique capable of manufacturing the stack with the chosen materials and whether it is feasible to employ in this case. Fortunately, silicon dioxide and silicon nitride are used in various applications for micro-electromechanical systems and microelectronics, and therefore the necessary deposition techniques are widely and commercially available. The dielectrics have been deposited through PECVD at PoliFab, Politecnico di Milano, and we obtained five manufactured stacks from the facility. 

For the future development of this prototype, we may consider operating the entire setup at cryogenic temperatures and as such, it is important that the materials chosen have low coefficients of thermal expansion (CTE). This further reinforced our choice of SiO$_2/$Si$_3$N$_4$ as the materials for the stack, being that they have thermal coefficients of $0.5 \times 10^{-6}\, \rm K^{-1}$ and $3.3 \times 10^{-6}\, \rm K^{-1}$~\cite{Tarraf_2003}, respectively. If we will operate the entire set up, including the stack, at cryogenic temperature, a dedicated study of the variation of the index of refraction as a function of temperature and of the stress induced by the CTE of the layers on top of each other may be necessary.

Once the materials are chosen, the other stack parameters must be decided upon, such as the thicknesses of the layers, the number of layers, the area of the stack, and the stack configuration (mirrored or not mirrored). While the area and the total stack thickness were constrained by experimental limitations, the other parameters were chosen using the optimisation procedure described in the next section. 

As we shall see, a mirrored chirped stack turned out to exhibit the highest boost factor. To make the substrate mirrored, we deposited a \SI{20}{nm} layer of Cr (for the sake of Au adhesion only) followed by \SI{75}{nm} of Au onto the substrate underneath the dielectric stack. The substrate is an ultraviolet grade fused silica plate of diameter \SI{50.08}{mm}, \SI{1}{mm} thickness, 2$\rm \lambda$ flatness and surface quality S/D 40/60. 

\subsubsection{\label{sec:sec4a2} Stack optimisation}
\begin{figure}[t]
	\begin{center}
	\includegraphics[width=0.45
	\textwidth]{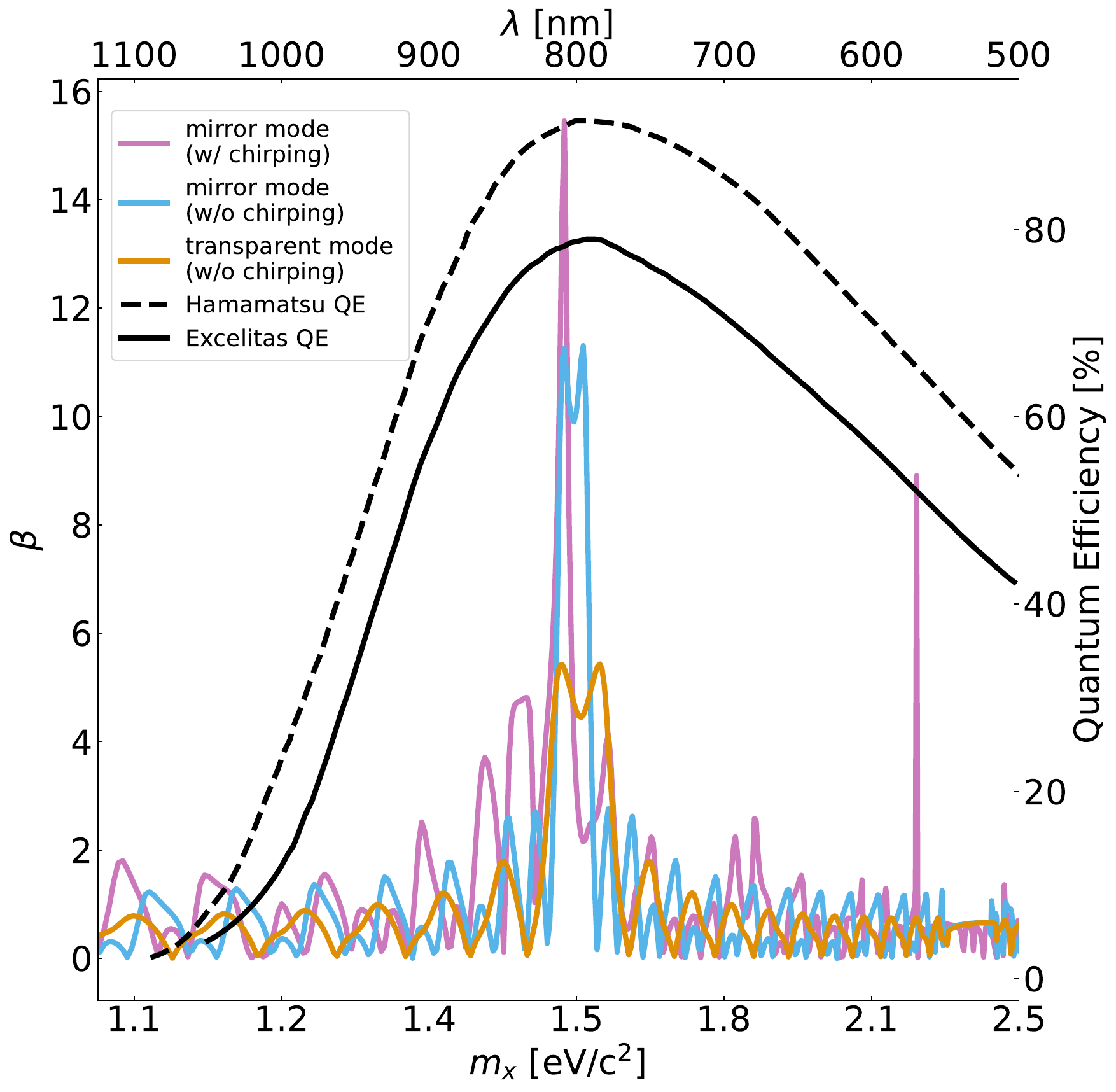}
	\caption{Boost factor spectra for different stack configurations. Each boost factor spectrum has been optimised to peak where the QE of the photosensor is at its maximum. 
    The dashed black curve refers to the first photosensor that was tested (Hamamatsu S12426-02 Si APD). The design of the stack was based upon this sensor, even though it was eventually not used in the final experiment. The solid black curve represents the QE of the Excelitas SPAD that was employed in the final configuration.}
	\label{fig:boost_factor_spectra}
	\end{center}
\end{figure}

When optimising the stack to maximise the boost factor, three parameters are of prime importance. First, the configuration of the stack (mirrored, chirped, etc.); second, the thicknesses of the layers; and third, the number of layers. The only constraints in the optimisation process are the materials (chosen beforehand), the total thickness of the stack (dictated by the PECVD machine limits) and its area. The goal of the optimisation is to ensure that the boost spectrum is maximised where the sensitivity of the photosensor is at its maximum. 

For this iteration, we used a mirrored stack since reflection of the $E$-field on the mirror enhances the boost value when compared against a transparent stack. The boost factor also benefited from the ``chirped'' thicknesses of the bilayers, that is, they were set such that they would increase linearly from the first bilayer to the last by a constant factor $\alpha$ (this was a free parameter in the optimisation algorithm), i.e. 
\begin{subequations}
\begin{align}
d_1 &= \alpha d_{n-1}\,,\\
d_2 &= \alpha d_{n}\,,
\end{align}
\end{subequations}
where $n$ is the total number of layers. For a comparison among the different configurations see Fig.~\ref{fig:boost_factor_spectra}.
We also repeatedly varied all thicknesses by a small amount (based on a Gaussian distribution centered around the nominal layer thickness with standard deviation of about \SI{40}{nm}) to further enhance the boost factor.

\subsubsection{\label{sec:sec4a3} Measurement of the layers} 
\begin{figure}[b]
	\begin{center}
	\includegraphics[width=0.47
	\textwidth]{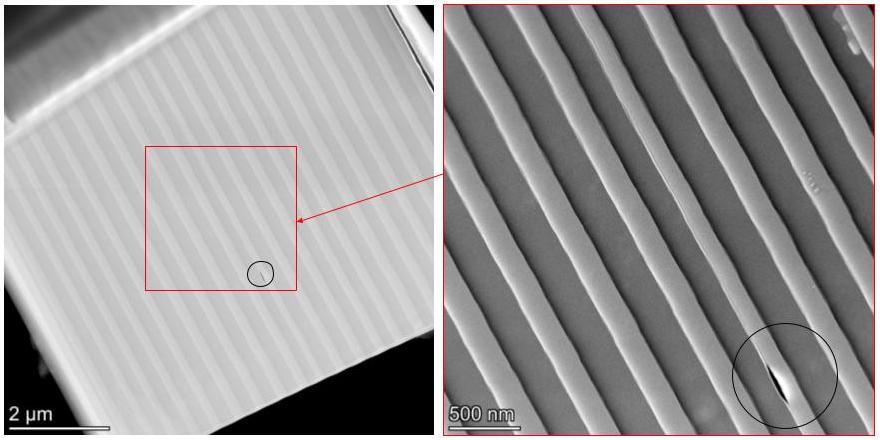}
	\caption{TEM image of one of the samples extracted via FIB, the scale is shown on the bottom left corner. The photo on the right shows a zoomed-in section (the red square pointed by the arrow) of the lamella on the left. The small crack (highlighted by a circle in both images) could be due to compressive stress during the deposition.}
	\label{fig:TEM}
	\end{center}
\end{figure}

Ideally, one would like to measure the thicknesses of the bilayers non-destructively and one way to achieve this is through spectroscopic ellipsometry (SE). This technique can indirectly measure thin film thickness, refractive index, surface roughness and much more by measuring the polarisation change as light reflects from the surface. In-situ SE allows real-time monitoring of thin film growth but not all chambers are equipped with the optical instruments required to perform this analysis--as was unfortunately the case for our stack. Ex-situ SE can be used to analyse single layers with ease, but the characterisation of multilayer stacks becomes more complicated. For multilayer stacks, a standard approach is to measure the complex refractive index of each material from reference single-layer samples and then use that information in the analysis of the SE data to constrain some of the free parameters in the fit. This procedure works well for stacks of 5--10 layers, but it becomes difficult and unreliable (due to the numerous parameters to fit) for stacks with more layers~\cite{hilfiker2019spectroscopic}, such as in our case. 
Thus, we opted for transmission electron microscopy (TEM), which we can easily access at the Analytical and Materials Characterization Core Technology Platform at NYUAD. TEM samples were also prepared at NYUAD via focused ion beam (FIB) sample preparation procedures. We milled several lamellae from two of the five stacks produced, welded them to an Omnigrid and thinned them to electron transparency (less than \SI{100}{nm}). Out of all the samples extracted only four turned out to give good TEM images, two of which belonged to the same stack used for data taking in the haloscope. 

As opposed to the model-based SE method, the TEM approach provides a direct measurement of film thicknesses. The main drawback is that it is a destructive technique, and consequently the TEM analysis can be performed on the actual stack used in the detector only when the experiment is over. Similarly to SE, TEM also gives a local characterisation of the multilayer stack such that multiple measurements would be needed to fully characterise the stack across the entire area. Although this is in principle possible, in practice it takes a lot of time and effort to extract and analyse more than a just a few samples. 

Figure~\ref{fig:TEM} shows the TEM image of one of the samples where one can count 43 out of the total 46 layers present (three of them were simply not extracted or not thinned enough to be seen via TEM). We also carried out an energy dispersive X-ray analysis (EDAX) for each sample to characterise the composition of the layers which demonstrated that there was good separation from one layer to another in terms of chemical composition.

Finally, we cooled down one of the samples to liquid nitrogen temperature to see how the stack would react to thermal shock. This last measurement was done in light of the future development of this experiment, where the haloscope could be fully operated in a cryogen-free dilution refrigerator with a transition edge sensor (TES) as the photodetector. The TEM analysis showed that the thermal stress did not cause any visible crack to the stack. This is even more reassuring being that in reality the stack would be cooled down gradually (although to even colder temperatures) and not as abruptly as we have done. 

\subsection{\label{sec:sec4b}{SPAD}} 

\begin{figure}[tb]
	\begin{center}
	\includegraphics[width=0.45\textwidth, trim=2cm 2cm 2cm 2cm]{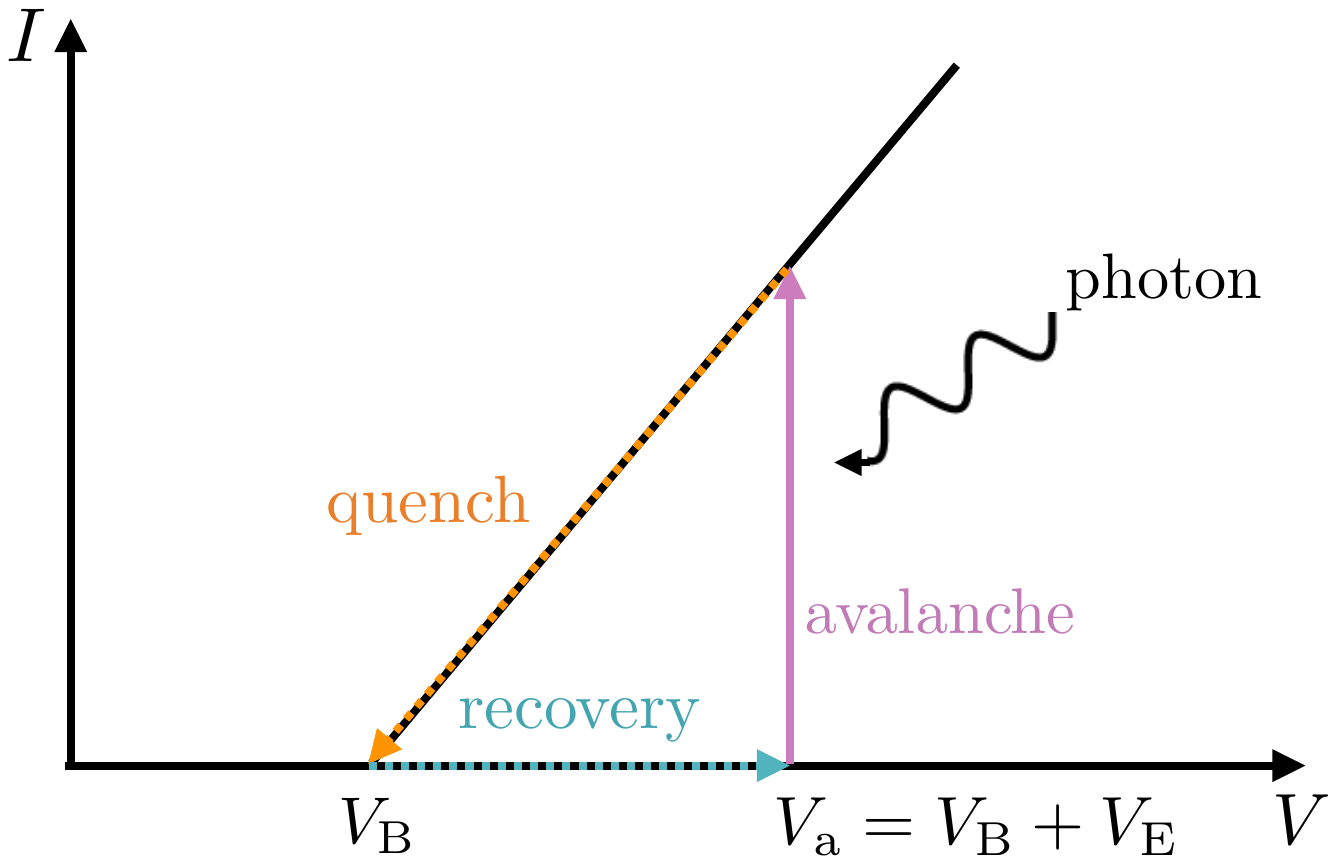}
	\caption{In its idle state, the SPAD is reverse-biased at a voltage $V_a$ above the breakdown voltage $V_B$ and no current is present. A photon impinging on the silicon can excite an electron from the valence band to the conduction band, leaving a hole behind. For the excitation process to happen, the photon must have a minimum energy, which depends on the band gap of the material (\SI{1.2}{eV} for silicon).}
	\label{fig:SPAD_mechanism}
	\end{center}
\end{figure}

A SPAD (also known as a Geiger-mode Avalanche Photodiode) is a p-n junction reverse-biased at a voltage $V_{\rm a}$, that exceeds the breakdown voltage $V_{\rm B}$ by an amount $V_{\rm E}$, called excess bias:
\begin{equation}
    \label{eq:avvoltage}
    V_{\rm a} = V_{\rm B} + V_{\rm E}\,.
\end{equation}
A photon absorbed in the silicon can generate an electron-hole pair. If injected in the so-called depletion (or multiplication) region, the carriers are accelerated by a strong electric field (greater than $3\times10^5$\,V/cm) such that even a single photo-generated carrier can trigger a diverging, self-sustaining avalanche multiplication of charge carriers by impact ionisation.

SPADs operate above the breakdown voltage, unlike ordinary APDs (avalanche photodiodes) which operate in reverse-bias below the breakdown voltage~\cite{COVAbookChap4}. Operation in this mode means that the current is proportional to the number of photons absorbed, although the limited gain (up to ${\approx}5\times 10^2$) does not allow for the detection of single photons~\cite{COVAbookChap4}.
Conversely, SPADs are sensitive to single photons but can only distinguish between zero photons or more than zero photons, i.e.~they are not photon-number resolving devices~\cite{Eisaman2011}.
Therefore, whilst an APD can be seen as a detector with an integrated amplifier, a SPAD can be considered a detector with a digital flip-flop~\cite{COVAbookChap4}.

Because of their ability to detect single photons, SPADs are to be preferred to APDs as photosensors in dielectric haloscopes. 
For  regular  APDs, removing any incident light is sufficient to immediately stop the multiplication process. However, in the case of SPADs, once triggered the avalanche will continue even in the absence of light.
To stop the cascade, the bias voltage has to be decreased to a value below the breakdown voltage through a high impedance path. This is achieved through a so-called quenching circuit. The role of this circuit can be seen in Fig.~\ref{fig:SPAD_mechanism}, which shows the typical operation mode of a SPAD.

The breakdown voltage of a SPAD strongly depends on the temperature of the p-n junction, with $V_{\rm B}$ decreasing with decreasing temperature~\cite{cova_1996}. Therefore, it is crucial to precisely stabilise the temperature during operation: if the temperature were to change while keeping $V_{\rm a}$ fixed, so too would the excess voltage $V_{\rm E}$ and $V_{\rm B}$ change. This is important as both the dark count rate (DCR) and the photodetection efficiency of the device are dependent on the excess voltage applied and consequently the dark count rate is dependent on the temperature. The degree to which the temperature needs to be stable depends on the specific sensor and is given by the slope of the ``voltage breakdown \emph{vs} temperature'' plot. The temperature stabilisation and sensitivity for our setup is discussed in Sec.~\ref{sec:sec5a}.

Dark counts refer to pulses present even in the absence of incident light and are generated by thermal effects. These counts represent the primary source of noise in the photosensor. Three effects contribute to such false detection events~\cite{Panglosse2020, COVAbookChap4, van_2010, cova_1996}:
\begin{itemize}
    \item The thermal release of charges. These can be generated within the depletion region, where they can immediately trigger an avalanche, or the undepleted region, where by diffusion they can get to the high-field region and initiate the cascade. Band-to-band thermal generation of carriers is rare due to the silicon band gap energy being \SI{1.2}{eV}. More common is the enhanced thermal emission of electrons through traps. In this process electrons rely on mid-gap trap energy levels that arise from impurities or crystalline defects to cross the potential barrier (trap-assisted thermal generation) or to tunnel through it (thermal generation via trap-assisted tunnelling) to the conduction band.
    \item Band-to-band tunneling and band-to-band trap assisted tunneling within the depletion region. At low temperature this is the major contribution to dark noise. 
    \item Secondary dark pulses caused by afterpulsing effects. During the avalanche some carriers can be trapped by deep levels\footnote{Transition-metal impurities are a common source of deep levels~\cite{COVAbookChap4}.} in the depletion region and released later on. Fabrication technology can greatly suppress afterpulsing.
\end{itemize}
To decrease the thermal contribution to the dark count rate, the SPAD can be cooled down by means of a Peltier cell. 

Another important parameter in characterising a SPAD is its photon detection efficiency, that is the ratio between the number of incident photons and the number of output current pulses~\cite{zappa2007principles}. The PDE is given by
\begin{equation}
    \text{PDE}(\lambda, V_{\rm E}) = {\rm QE(\lambda)} \times P_{\rm b}(V_{\rm E}, T)\,,
\end{equation}
where $\rm QE$ is the quantum efficiency\footnote{One can measure the quantum efficiency using a calibrated light source with the SPAD operated in linear mode.} and $P_{\rm b}$ is the probability that the primary photogenerated electron-hole pair initiates an avalanche. 
While the quantum efficiency comprises the dependency on the wavelength of the incident photon ($\lambda$), $P_{\rm b}$ carries the dependency on the excess voltage $V_{\rm E}$ and the temperature $T$\footnote{The dependence on the temperature is very mild above \SI{800}{\nm}.} at which the SPAD is operated.

The $\rm QE$ takes into account the probability that the photon crosses the anti-reflection layer and then creates an electron-hole pair in the sensitive region.

\begin{figure}[tb]
	\begin{center}
	\includegraphics[width=0.45\textwidth, trim=0cm 0cm 0cm 0cm]{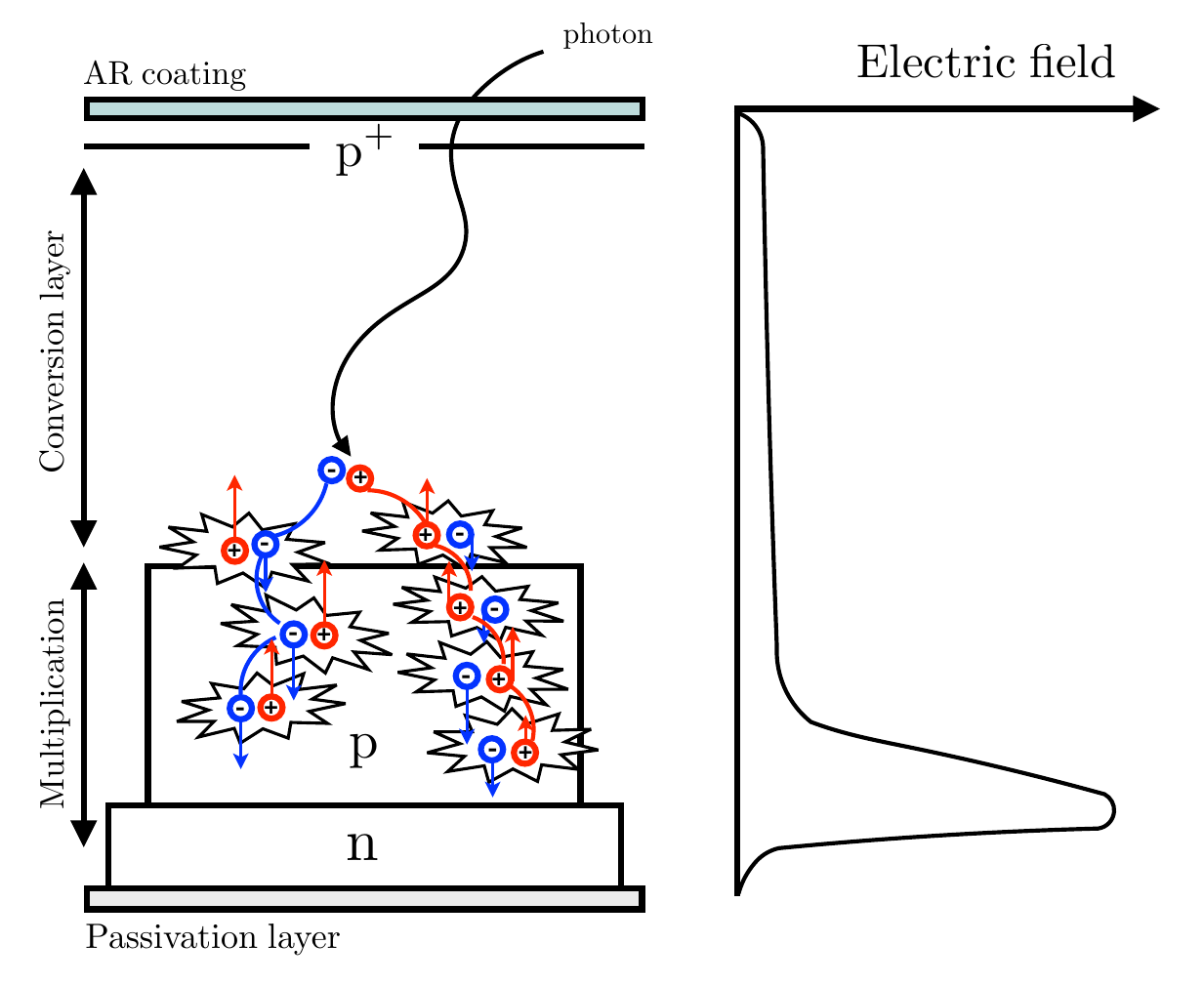}
	\caption{The figure shows schematically the structure of the Excelitas SPAD (the layer thicknesses are not in scale). The photon has to cross the anti-reflection (AR) coating before being absorbed in the silicon. The low field ``conversion layer'' converts the photon into an electron-hole pair. In the ``multiplication layer'' the pair may trigger a diverging avalanche to produce a detectable electric pulse. A passivation layer (oxide layer) protects the silicon from moisture and other contaminants where the electric field is the highest.}
	\label{fig:p-n_junction}
	\end{center}
\end{figure}

$\rm P_{\rm b}$ depends on all that happens after.
To undergo multiplication, the carrier will then have to diffuse into the depletion region (unless generated there) without recombining and once in the depletion region, the carrier may trigger an avalanche (see Fig.~\ref{fig:p-n_junction}).
The diffusion of holes and electrons depends on their differing mobility inside the lattice, their lifetime (which can vary a lot depending on the purity of the silicon), and the temperature of the junction. 
Finally, the chance for a photon to trigger a cascade increases with increasing excess voltage at the expense of a higher dark count rate.
For more details about photon detection efficiency in SPADs, we suggest consulting Refs.~\cite{panglosse2021modeling, xu2016new, hosseinzadeh2017analysis, mazzillo2008quantum, zappa2007principles, COVAbookChap4, cova_1996}. 

\subsection{\label{sec:sec4c}{Assembly}}
\begin{figure}[t]
	\begin{center}
	\includegraphics[width=0.45
	\textwidth]{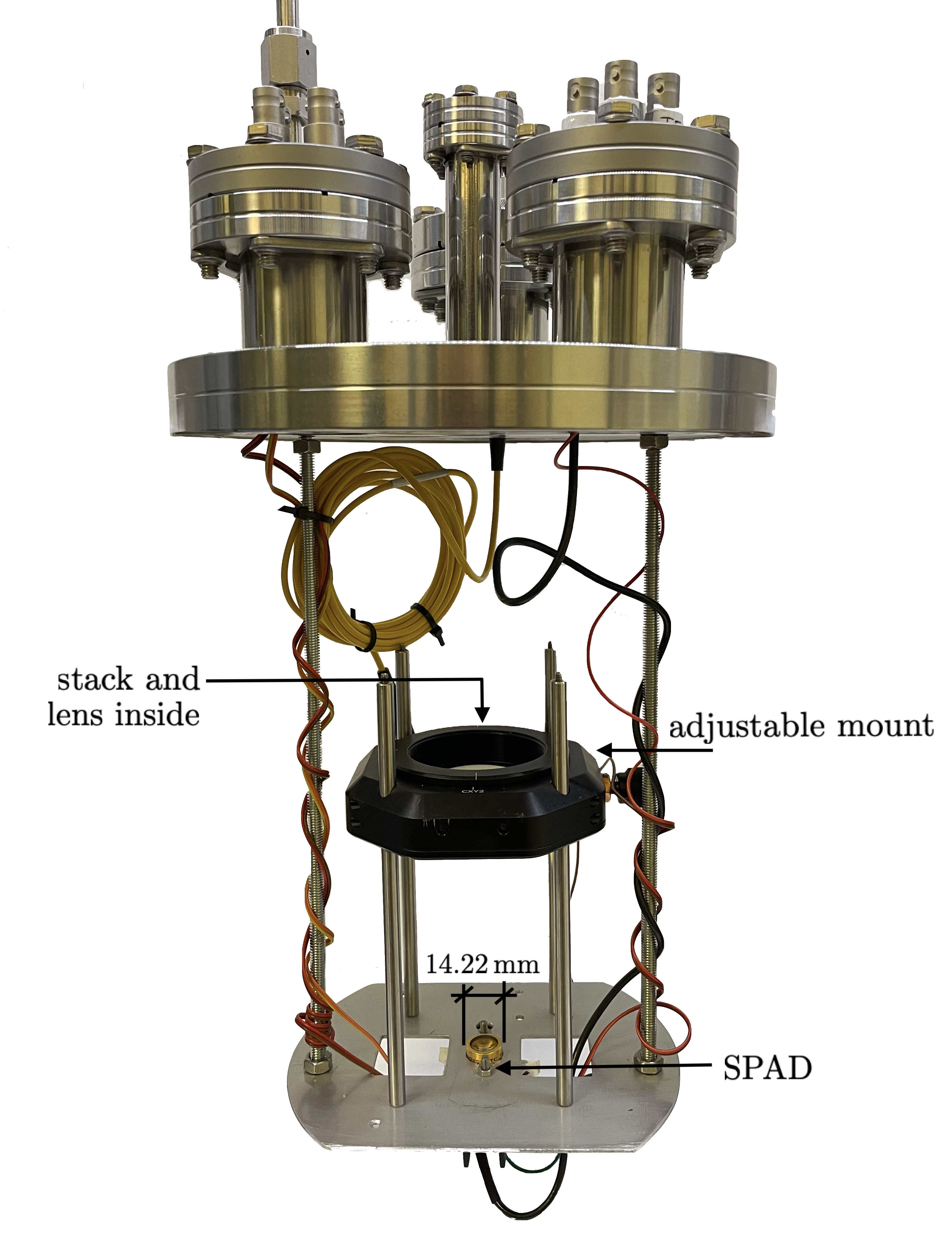}
	\caption{Photo of the MuDHI dielectric haloscope hanging from the lid of the stainless steel chamber (not shown).}
	\label{fig:haloscope}
	\end{center}
\end{figure}
\begin{figure}[t]
	\begin{center}
	\includegraphics[width=0.49\textwidth]
	{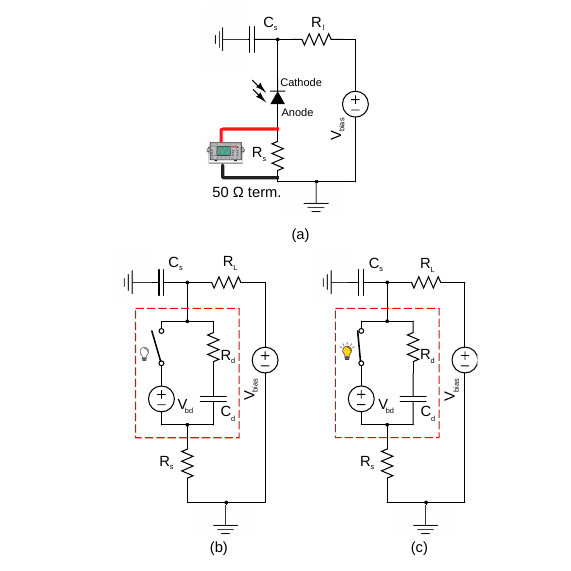}
	\caption{(a) PQC employed for the SPAD readout. (b) Equivalent circuit for the SPAD in absence of light. Switch is open and the junction capacitance is charged to bias. (c) Equivalent circuit for the SPAD after photons have hit the sensor. The switch is closed and the junction capacitance discharges until it reaches the breakdown voltage}
	\label{fig:spad-circuit-haloscope}
	\end{center}
\end{figure}
The experimental assembly consists of a Thorlabs optical cage mounted on a custom designed base plate. The optical cage houses the dielectric stack on top of the aspherical lens in an adjustable mount (featuring submillimeter precision knobs for positioning along the x-y plane, and four rods onto which the cage is mounted and can slide freely for positioning along the z-axis). The SPAD is fixed on the base plate such that it lies at the focus of the aspherical lens and in order to ensure proper alignment, the lens position is adjusted in the x-y plane using a red laser diode coupled to a collimator. The collimator is inserted in a Thorlabs cage plate adapter mounted directly on top of the lens holder, such that the light is perfectly perpendicular to the lens. The alignment of the lens along the z-axis is not only achieved by focusing the aforementioned beam from the laser diode, but also by focusing a distant object onto the base upon which the photosensor is mounted. To verify that the lens is aligned properly with the SPAD, we shine ambient light onto the SPAD through an optical feedthrough and a fibre connected to the cage plate adapter. Upon removal of the fibre, the SPAD does not see any light, thus verifying its proper alignment. 

The Excelitas SPAD module is equipped with a thermoelectric cooler (right underneath the SPAD) and a thermistor to allow temperature regulation. This entire setup, shown in Fig.~\ref{fig:haloscope}, is placed inside a light tight stainless steel chamber to prevent light leakage. 

The SPAD readout is achieved using a passive-quenching circuit (PQC). As shown in Fig.~\ref{fig:spad-circuit-haloscope}, the key components of this circuit are the ballast resistor $R_{\rm L}$, a stray capacitance $C_{\rm S}$ (i.e.~a capacitance to ground connected by $R_{\rm L}$ to the positive diode terminal--the cathode) and a sense resistor $R_{\rm S}$ to enable voltage readout on the oscilloscope (this is in series with the \SI{50}{\Omega} oscilloscope termination). To understand the PQC, we can model the SPAD as an RC circuit with diode resistance $R_{\rm d}$ and junction capacitance $C_{\rm d}$. This junction capacitance is normally charged to the bias voltage $V_{\rm a}$. When light impinges on the sensor and the bias voltage is above the breakdown voltage, the avalanche is initiated and the avalanche pulse is stored in $C_{\rm d}$. The avalanche triggering corresponds to the switch in the diode equivalent circuit being closed. At that point $C_{\rm d}$ discharges down to the breakdown voltage $V_{\rm B}$ for the duration of the quench time, which is set by $R_{\rm L}$ and $R_{\rm d}$ in parallel, and the total capacitance $C_{\rm d}+C_{\rm s}$~\cite{COVAbookChap4}. 
At this point, the switch opens again and the junction capacitance is exponentially recharged up to the bias voltage with time constant $T_{\rm r} = R_{\rm L}(C_{\rm s}+C_{\rm d})$. At the end of the recovery period $T_{\rm r}$, the SPAD is again ready for detection. 

\subsection{\label{sec:sec4d}{DAQ system}}
\begin{figure*}[t]
\centering
\subfigure[]{
\label{fig:DCRvstemp}
\includegraphics[width=0.3\textwidth, trim=1cm 1cm 1cm 0.98cm]
{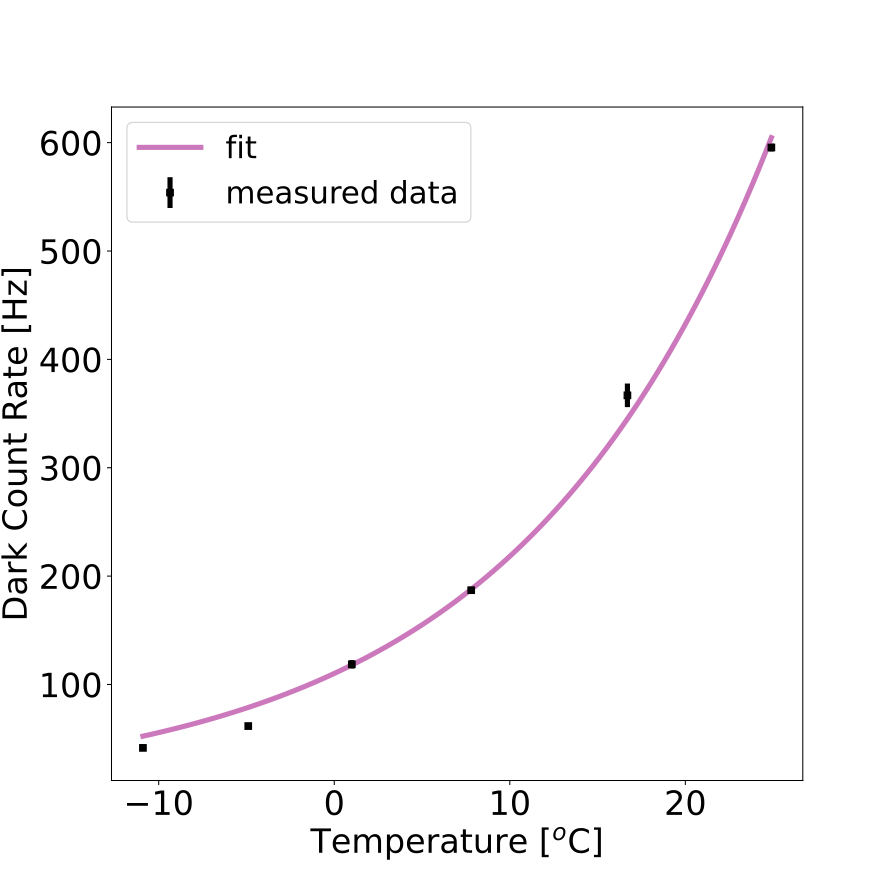}}%
\hfill
\subfigure[]{
\label{fig:DCRvsVe}
\includegraphics[width=0.3\textwidth, trim=1cm 1cm 1.55cm 1cm]{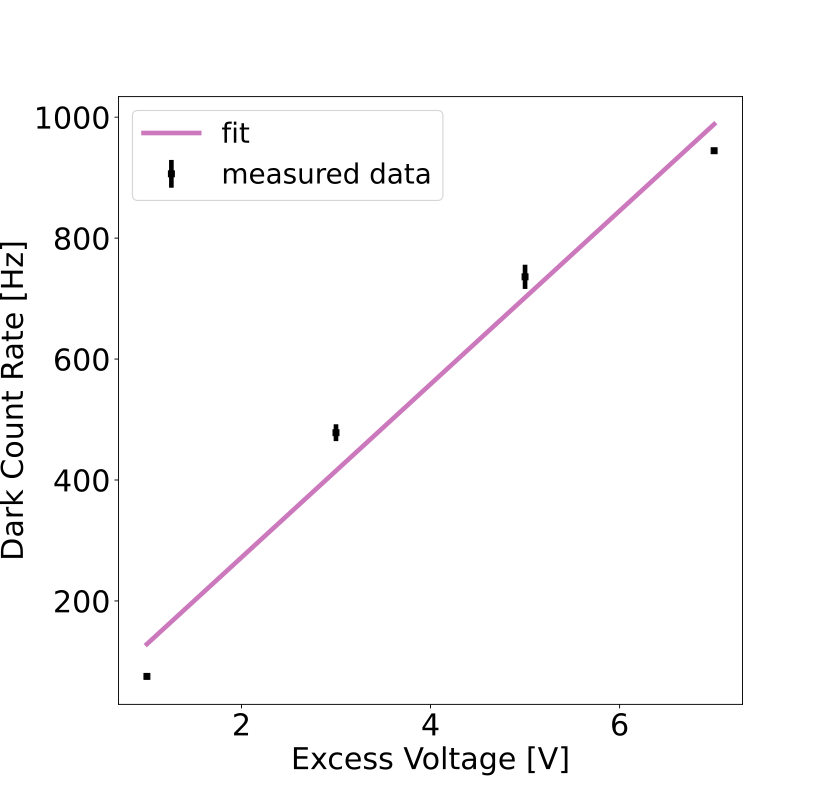}}%
\hfill
\subfigure[]{
\label{fig:Vbd-temp}
\includegraphics[width=0.3\textwidth, trim=1cm 1cm 1cm 1cm]{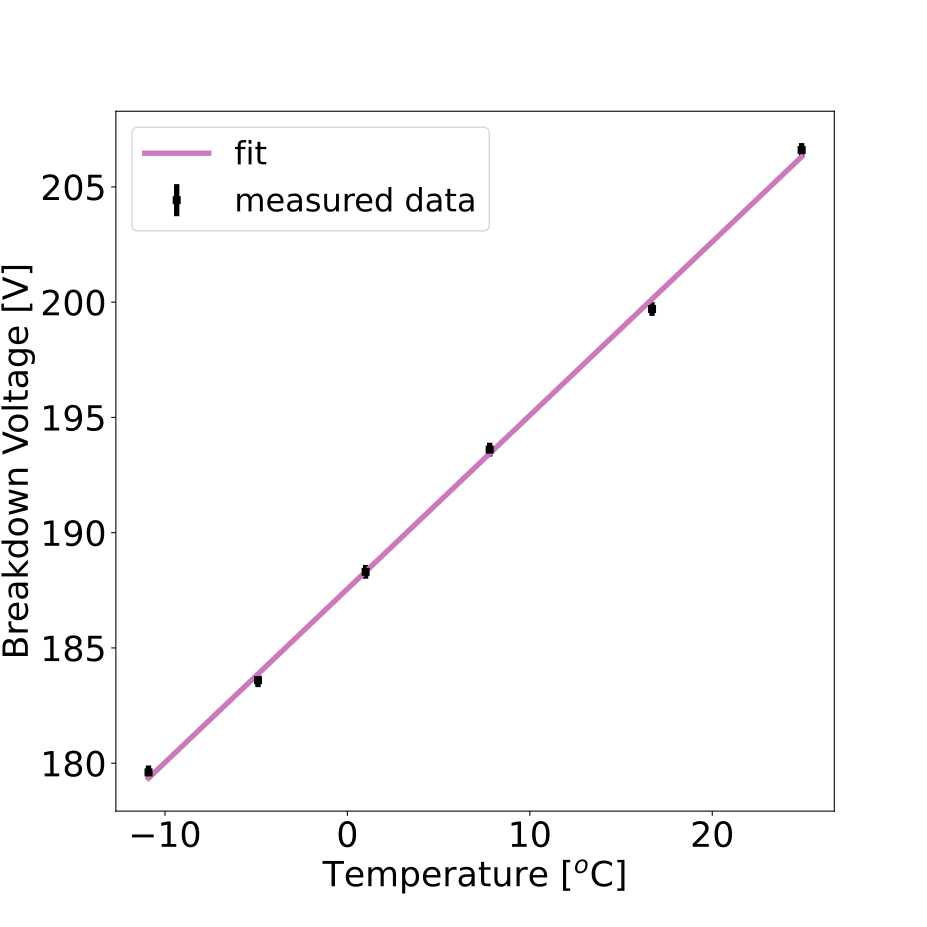}}
\caption{(a) Dark count rate as a function of temperature at \SI{1}{V} excess voltage. The dependence of dark-count rate on temperature is exponential. (b) Dark count rate as a function of excess voltage at the constant temperature of \SI{-5}{^oC}. (c) One way to check that the SPAD is working as expected is to verify that the breakdown voltage decreases as the sensor cools down. Changes in temperature of \SI{0.1}{^oC} do not effect the breakdown point of the device. The temperature coefficient of breakdown voltage for constant gain is \SI{0.75}{V/\degreeCelsius}, in good agreement with the range of 0.5--\SI{0.9}{V/\degreeCelsius} reported by the manufacturer (see Ref.~\cite{Excelitas_datasheet}).}
\end{figure*}
The signal and temperature outputs from the SPAD are fed into the data acquisition (DAQ) system, which allows us to measure the SPAD counts and to implement a proportional–integral–derivative (PID) controller in order to regulate the temperature. 

The DAQ consists of Nuclear Instrumentation Module (NIM) standard devices. The N625 Quad Linear Fan In/Fan out module is used for inverting and copying signals where necessary. The SPAD output signal has to be inverted since the output of our circuit gives a positive edge pulse but the NIM standard works on negative edge pulses. A copy of the inverted signal goes into the N844 leading edge type discriminator (LTD). This discriminator provides a NIM logic high when the signal is below a set threshold (recall that signals are negative). The threshold in this experiment is set to \SI{-5}{mV} which is sufficiently low to reject the baseline noise. Since the amplitude of dark pulses (i.e.~pulses generated when no light is incident) is on the order of \SI{30}{mV}, the signal loss is completely negligible. The output of the discriminator goes to the N93B dual timer module and then into the N1145 Quad Scaler and Preset Counter Timer module, which counts the number of NIM logic high pulses within a given time period. 

To implement the PID controller, a Keithley multimeter is used to read the temperature from the SPAD thermistor. A PID script written using the PyVisa library is implemented to regulate the PID settings and record the temperature data for each run.  
\section{\label{sec:sec5} Results}
\subsection{\label{sec:sec5a} SPAD characterisation}

As mentioned in Sec.~\ref{sec:sec2}, three SPADs were tested for this experiment. Since the data from the Excelitas SPAD was used to make the final exclusion curve, we only present the characterisation of this SPAD. The Hamamatsu device had extremely high dark count rates and did not show the characteristic SPAD behaviour in Geiger mode. The Laser Components SPAD did not come with an in-case TEC, and using an external Peltier cell did not prove to be a reliable method of cooling due to temperature fluctuations of $\sim \mathcal{O}(1^\circ \rm C)$. 

The dependence of the dark count rate (DCR) on temperature for the Excelitas SPAD is shown in Fig.~\ref{fig:DCRvstemp}. Ideally, to keep the DCR to a minimum, one would use the lowest temperature when operating the SPAD. However, keeping a stable temperature below \SI{-5}{^oC} for time periods on the order of hours proved challenging. Therefore, the SPAD was run at \SI{-5}{^oC} and \SI{1}{V} excess voltage, values that allowed for consistent and stable operation. We were able to achieve temperature stability within \SI{0.1}{^oC} using the PID control setup, which is a low enough margin so as to not introduce errors in the estimated breakdown voltage as shown in Fig.~\ref{fig:Vbd-temp}. The choice to operate the SPAD at \SI{1}{V} above the breakdown voltage was made to ensure a high PDE, which as mentioned increases with increasing $V_{\rm E}$, while still maintaining a low DCR. The dependency of the DCR on the excess voltage at \SI{-5}{^oC} is shown in Fig.~\ref{fig:DCRvsVe}.

During measurements of the DCR, we observed a peculiar behavior wherein the DCR was seen to change despite keeping the temperature and the excess voltage constant. It was noted that when this happened the amplitude of the signal got larger over time, which allowed more pulses to pass the threshold, thereby increasing the measured rate. A change in the signal amplitude could be attributed to a change in the breakdown voltage despite the temperature being constant. It was indeed confirmed that any time the dark count rate deviated from its nominal value, so too did the breakdown voltage, despite the temperature not changing. The manufacturer later indicated that the change in DCR may be linked to a bistability or a multistability effect, that is well-known in SiAPDs operating in Geiger Mode. Some suggest this is related to the density and distribution of defects in the sensor lattice~\cite{bistability, buchinger1995identification}. To remedy this defect, we ensure that the amplitude of the pulses remains constant throughout the duration of the data acquisition\footnote{The oscilloscope is sensitive to variations in the amplitude within \SI{1}{mV}, while instabilities in the sensor cause amplitude deviations of \SI{4}{mV} or more.} by monitoring the pulses on the oscilloscope, discarding the data that exhibited such anomalous behaviour.



\subsection{\label{sec:sec5b} Observed upper limits}

The experiment was conducted into two phases: what we call the ``off measurement'', where counts are recorded from the haloscope without the stack ($N_{\rm off}$); and the ``on measurement'', where we operate the haloscope together with its stack ($N_{\rm on}$). 
We compute $90\%$ confidence level upper limits using the profiled log-likelihood ratio method, described in appendix~\ref{app:statistics}.

The measurements were taken at \SI{-5}{^oC} and \SI{1}{V} excess voltage. The ``on measurement'' ($T_{\rm on}$) lasted two hours, while the ``off measurement'' lasted 30 minutes ($T_{\rm off}$). As mentioned in Sec.~\ref{sec:sec3b} the longer the exposure during the ``on measurement'', the greater the the signal to noise ratio. Unfortunately, the data collection time was limited by the previously discussed anomaly wherein the breakdown voltage would start changing. To ensure that the breakdown voltage did not change while measurements were taken, the pulse counting was split into runs of ten minutes each and the breakdown voltage was measured in between. Runs where the breakdown voltage changed from its nominal value were discarded. The final observed count rates were $n_{\rm on} = N_{\rm on}/T_{\rm on} = 98.6\,\rm Hz \pm 2.6\,\rm Hz $ and $n_{\rm off} = N_{\rm off}/T_{\rm off} = 96.5\,\rm Hz \pm 2.3\,\rm Hz$, consistent with no signal observed. The corresponding $90\%$ upper limit on the median observed dataset is $2.3\,\rm Hz$.

\begin{figure}[tb]
    \raggedleft
    \includegraphics[width=0.46\textwidth, trim= 1cm 0cm 0ccm 0cm]{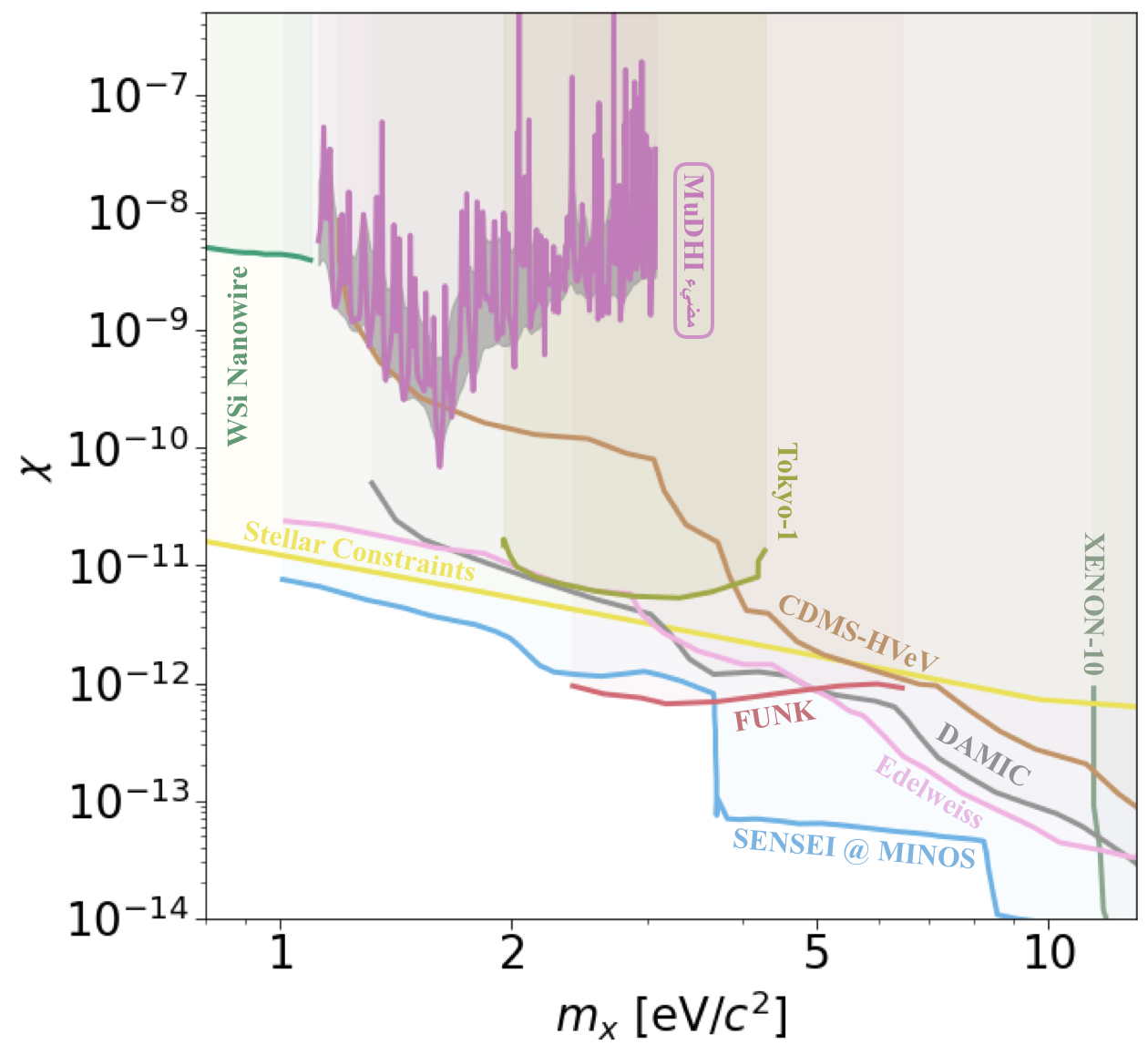}
    \caption{Current constraints on the kinetic mixing parameter $\chi$ as a function of the dark ph
oton mass $m_X$. In magenta the result from MuDHI. The other experiments included are stellar constraints from Ref.~\cite{An_2013_stellar}, direct detection constraints from Xenon10~\cite{An_2015_xe10}, CDMS~\cite{Agnese_2018_CDMS}, DAMIC~\cite{Aguilar_Arevalo_2019_damic}, 
    Edeleweiss~\cite{Arnaud_2020_edelweiss},
    SENSEI~\cite{Barak_2020_sensei},
    Tokyo-1~\cite{Suzuki_2015_Tokyo}, and
    FUNK~\cite{Andrianavalomahefa_2020_funk} collaborations. We also include constraints from Ref.~\cite{Hochberg_2019_nano} that uses WSi nanowires as target for dark photon detection.}
    \label{fig:upperlimits}
\end{figure}

Figure~\ref{fig:upperlimits} shows the observed upper limit at 90\% CL of the kinetic mixing parameter $\chi$ as a function of the dark photon energy using the measured boost factor. The upper limits using the 15$^{\rm th}$ and 85$^{\rm th}$ percentiles of the boost factor spectrum are also shown as a semi-opaque black band. The same plot shows the most updated constraints using cosmological, experimental, and astrophysical bounds. Since the thickness of the layers in the manufactured stack differed from the nominal thicknesses (i.e.~the theoretical ones used during the PECVD process), the measured thicknesses were used for constructing the observed upper limit at 90\% confidence level. The measurement of the thickness of the layers carries with it some error, such that an error estimation on the boost factor is also required. 
For most of the layers we managed to take four distinct thickness measurements, using four different TEM samples. Two samples were taken from the stack used in the actual dark photon experiment and two from another stack--both of which were fabricated in the same PECVD run at PoliFab. 
The designed boost factor differs from the real boost factor due to uncertainties in the PECVD manufacturing process. Additionally, the measured boost factor differs from the real boost factor due to measurement uncertainty. However, even though the measured boost factor is not a perfect proxy for the real one, it does provide a good representation for the quality of real boost factors that are producible through PECVD relative to the designed boost factors we are attempting to create.
When creating limits from our real stack, simply considering the uncertainty per mass point is a simplification of the true correlated uncertainty across masses, and it is very challenging to implement this as a nuisance parameter in the inference. Therefore, for this result we indicate an estimated envelope of the error due to measurement error of the thicknesses together with the limit using the boost factor based on measured thicknesses. 

\begin{figure}[tb]
    \centering
    \includegraphics[width=0.45\textwidth]{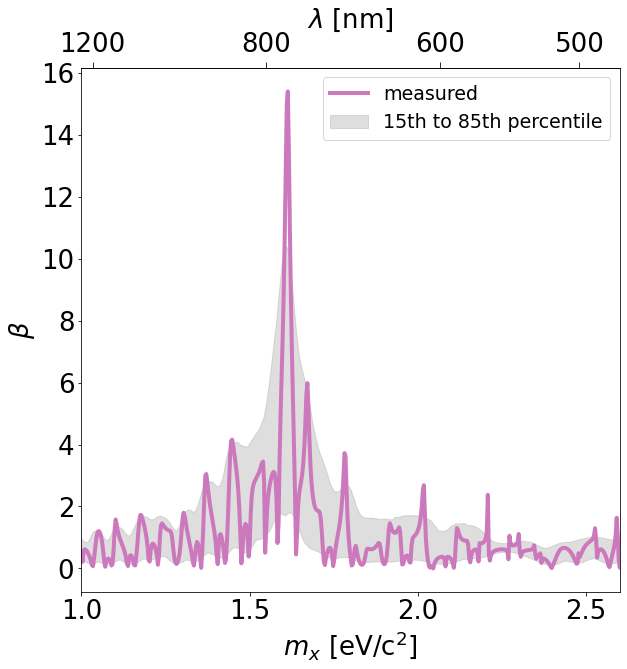}
    \caption{Boost factor spectrum as a function of the dark photon wavelength and energy. The shaded band shows the  15$^{\rm th}$ to 85$^{\rm th}$ percentile of boost factors at each mass. The variations are due to the constructive and destructive interference of light in the stack and depend on the thickness and refractive index of the layers.}
    \label{fig:boostspectrumbands}
\end{figure}

To estimate the error on the boost factor given the uncertainty of each layer width,
we perform a series of toy simulations of the stack with thicknesses drawn from Gaussians with mean and standard deviation corresponding to the measured thickness and uncertainty. The 15$^{\rm th}$ to 85$^{\rm th}$ percentile of the resulting boost factor is shown as a grey band in Fig.~\ref{fig:boostspectrumbands}, and the limit corresponding to this envelope is included in Fig.~\ref{fig:upperlimits}. The maximum boost factor in the simulation varies between ${\approx}5$ to ${\approx}15$, with around $97\%$ of boost factor peaks in the region 1.4--$\SI{1.8}{\eV/c^2}$.

As the error in the measurement of the layer thicknesses is the dominant source of uncertainty (compared to other sub-leading uncertainty components such as the error on the PDE--known within a few percent--or the error on the lens alignment), it may be preferable to optimise future stacks not for maximal boost factor, but rather for robustness against error as suggested in Ref.~\cite{AlvarezMelcon:2020vee}.

\section{\label{sec:sec6} Conclusion and future outlook}
We have reported on the design, construction, characterisation, operation, data analysis, and results of MuDHI. This prototype multilayer dielectric haloscope used an affordable and commercially available photosensor, a SPAD, for the detection of dark photons. For the first time, the feasibility of making and operating a dielectric stack which is both chirped and with a high number of layers has been assessed. An algorithm was developed to determine the number of layers, the chirping factor, and the individual thicknesses in order to optimise the theoretical performance of the stack. The 23 bilayers of SiO$_2$ and Si$_3$N$_4$ were sectioned with FIB  and subsequently analysed with TEM. These measured thicknesses together with their uncertainties were used to evaluate the real boost spectrum of the stack. 
A cryogenic TEM evaluation of the stack was performed by immersing the lamella in liquid nitrogen, and no evident damage was observed. This result will be useful if the entire haloscope will be operated at cryogenic temperatures in later developments of the experiment. 

The experiment was carried out into two steps, one with the stack (``on measurement''), and one without the stack  (``off measurement'').
With a two hour ``on measurement'', and after correcting for the photodetection efficiency of the SPAD and the collection efficiency at the focal point obtained from the GEANT4 simulation, no signal excess was observed.  As such, constraints were placed on the kinetic mixing coupling constant between dark photons and ordinary photons at 90\% confidence level in the mass region where the detector is sensitive, with a minimum of $6.86\times 10^{-11}$ reached at \SI{1.61}{eV/c^2}. 

Due to the disparity between the designed and the measured thicknesses, the actual boost spectrum differs quite significantly from that which was derived theoretically. For this reason, we will work on improving the robustness between the designed and measured boost factor for the next iteration of the stack, rather than aiming only at its maximisation. To achieve this, we plan on employing a substrate with higher flatness and better scratch-dig specification and using a PECVD machine equipped for in-situ SE for manufacturing the stack. We also plan on using ex-situ SE (as outlined in Ref.~\cite{hilfiker2019spectroscopic}) and TEM after the deposition to compare the two methodologies. Another parameter that we foresee to measure is the actual reflectivity of the mirror layer, which can then be implemented in the simulation. Further, possible systematic effects due to three dimensional effects and imperfections will be studied to better characterise the response of the stack. 

Given the instability of the SPAD and its relatively high dark count rate, the second stage of the experiment will see the implementation of a superconducting transition-edge sensor that will be developed by collaborators at INRiM (Istituto Nazionale di Ricerca Metrologica, Torino, Italy). The transition temperature will be kept under \SI{100}{mK} in order to obtain a good energy resolution. The lower dark count rate (expected to be around $10^{-4}$\,Hz) and the capability of the TES to distinguish one photon from two or more (which could, for example, arise from muon interaction within the TES) will allow the experiment to probe areas of parameter space with weaker mixing couplings. Everything staying the same except for the dark count rate (i.e.~boost factor spectrum, quantum efficiency, collection efficiency, and exposure time same as in MuDHI) we expect the sensitivity to improve by about 1000 at the boost factor peak. Given the TES will be able to operate more stably, we also predict the acquisition time to be of the order of $\mathcal{O}$(10 days) in a cryogen-free dilution refrigerator, so that the 1000 factor improvement in sensitivity is an extremely conservative estimation.

\begin{acknowledgments}
Indispensable in triggering our interest for the detection of dark photons using dielectric haloscopes was the initial contact with Dr Ken Van Tilburg. Fundamental to the development of the detector were Dr Masha Baryakhtar, Dr Junwu Huang, and Dr Robert Lasenby, whose paper ``Axion and hidden photon dark matter detection with multilayer optical haloscopes''~\cite{Baryakhtar_2018} has been very helpful in the first stage of this work. 
We are grateful to Dr Alessia Allevi and Dr Maria Bondani for the fruitful conversation about single-photon detection sensors back in November 2018. We thank Jean-Matthieu Bromont, sales engineer at Hi-Tech Detection Systems partner of Excelitas Technologies, who was always available for support and advice throughout the project. We sincerely thank Jens Krause and Philippe Bérard from Excelitas Technologies for their guidance and suggestions. 
For the stack manufacturing, we thank Dr Andrea Scaccabarozzi and Dr Claudio Somaschini at PoliFab. For the statistical analysis, we express our gratitude to Dr Giacomo Vianello, who, before leaving academia for a successful career in data science, passed on his knowledge on the significance of an excess in counting experiments. We thank Jinumon Govindan, our workshop technician at NYUAD, who skillfully built the mechanical parts needed for the experiment.
This work was supported by the NYUAD Research Enhancement Fund. Alex Millar is supported by the European Research
Council under Grant No. 742104 and by the Swedish Research Council (VR) under Dnr
2019-02337 ``Detecting Axion Dark Matter In The Sky And In The Lab (AxionDM).'' 
Knut Dundas Morå is supported by the National Science Foundation under Award no. 1719286.
Last but not least, this work was made possible thanks to the heterogeneous expertise of its authors and their willingness to share their knowledge. As is often the case, effective collaboration and team-work are the cornerstones of scientific achievement. 
\end{acknowledgments}


\appendix
\section{Derivation of the expected rate formula}
In a dielectric haloscope the power is amplified as the square of the boost factor (see Eq.~\eqref{eq:dhflux})
\begin{equation}
    \frac{P_{\gamma}}{S} = \frac{\beta^2E_0^2}{2}\eta\,,
\end{equation}
where $\eta$ is introduced to account for the efficiency of the detector. This includes the PDE at the specific wavelength considered\footnote{The PDE as a function of wavelength at \SI{1}{V} excess voltage was provided by the manufacturer. Alternatively, one can obtain the PDE as a function of wavelength by scaling the quantum efficiency spectrum knowing that the PDE at \SI{1}{V} excess voltage at \SI{830}{nm} is 2.1\%. Both pieces of information are given by the manufacturer's datasheet~\cite{Excelitas_datasheet}.} and the collection efficiency at the focal point obtained from the GEANT4 simulation ($0.71.8\%$ in our case). 
Using Eq.~\eqref{eq:expectedrate}, the dark photon flux then reads
\begin{equation}
    \Phi = \frac{P_{\gamma}}{A\omega} = \frac{5.2}{\text{day}\;\text{cm}^2}\frac{10^{23}\text{eV}}{m_X}\chi^2\beta^2\eta \,,
    \label{eq:rate-dp}
\end{equation}
where we have assumed a DM density of 0.3\,GeV/cm$^3$.
When constructing upper limits, we get the number of counts $N$ compatible with the null hypothesis and then calculate the $\chi$ value excluded for a given $N$ at each dark photon energy. So,
recalling that the flux is a number of events per unit time per unit area 
\begin{align*}
    \Phi &= \frac{N}{\,A\,t}\,,
\end{align*}
we can get an expression for the dark photon kinetic coupling constant
\begin{equation}
    \chi = \sqrt{\frac{N m_X}{ 5.2 \times 10^{23} \, A \, t \times \beta^2 \times \eta}\,\left(\text{days cm}^2\text{eV}^{-1}\right)}.
\end{equation}

\section{\texorpdfstring{$\langle\cos ^2 \theta\rangle$}{} derivation}
\label{appendix:cos-derivation}
For a given function $f(x)$, the average over a continuous interval is given by
\begin{align*}
    \langle f(x) \rangle &= \frac{1}{b-a}\int_{a}^{b}f(x)\,dx
\end{align*}

Over a spherical surface, we can similarly write for $\cos^2 \theta$
\begin{align*}
    \langle \cos^2 \theta \rangle &= \frac{1}{A}\int_{A}\cos^2{\theta}\,dA
\end{align*}
Recall that $\theta$ is the angle between the polarisation $\hat X_0$ of the dark photon and the plane of the interface, i.e.~an angle between a line and a plane, while $\theta'$ is the angle between the polarisation vector and the vector normal to the surface.

The infinitesimal surface area is $dA = r^2 \sin\theta'\,d\theta'\,d\phi$ and the surface area of the sphere is $A=4\pi r^2$.  It can be verified that
\begin{align*}
    \langle \cos^2 \theta \rangle &= 1 - \langle \cos^2 \theta' \rangle\,
\end{align*}
thus
\begin{align*}
    \langle \cos^2 \theta \rangle &= 1 - \frac{1}{4\pi {r^2}}\int_{0}^{2\pi}\int_{0}^{\pi} \cos^2\theta' \, {r^2} \, \sin\theta' \,d\theta' \, d\phi\\
    &= 1 - \frac{1}{4\pi} \times 2\pi \times \frac{2}{3} \\
    &= \frac{2}{3}
\end{align*}

\section{GEANT4 simulation}
\label{appendix:GEANT4}

In order to enhance the detection probability, the shape, type, and placement of the lens was optimised to maximise the fraction of photons that hit the photosensor area placed at various focal distances. Specifically, we conducted the optimisation based on geometric optics followed by a GEANT4 simulation in order to calculate more precisely the collection efficiency at the focal point, while accounting for surface imperfections, aberrations, and light attenuation. 

\subsection{Geometric Optimisation}
To optimise the shape of the lens, we model it mathematically in 2D using cubic splines. Specifically, we define a finite, strictly increasing sequence of points $S = \{x_i\} \subset [0,1]$ where $\{0,1\} \in S$ and then connect each consecutive pair of points $x_i,x_{i+1}$ with cubic polynomials $S_i(x)$ such that $S_i(x_i) = x_i$ and $S_i(x_{i+1}) = x_{i+1}$. The system is then solved to obtain a smooth curve $s(x): [0,1] \to \mathbb{R}$ defined like so:
\begin{align*}
    s(x): [0,1] &\to \mathbb{R}\\
    x &\mapsto S_i(x) \text{ if } x \in [x_i,x_{i+1}]\,.
\end{align*}
A vector equation for the lens is created by applying a translation vector $\vec{T}$, a rotation matrix $R(\theta)$, and a scaling factor $\chi$ to the vector
\begin{equation*}
    \vec{s}(t) = \begin{pmatrix}
    t \\ s(t)
    \end{pmatrix},
\end{equation*}
in order to create the following parametric equation for the lens' surface:
\begin{equation*}
    \vec{S}(t) = \chi R(\theta)\, \vec{s}(t) + \vec{T}\,.
\end{equation*}

With this formulation it is possible to obtain the Frenet-Serret frame (i.e.~tangent $\hat{T}(t)$, normal $\hat{N}(t)$, binormal $\hat{B}(t)$ of the curve above) and thus perform a geometric propagation of rays through the parametrised lens surface $\vec{S}(t)$. Specifically, assuming an incoming ray in the direction $\hat{k}$ we can predict the change of direction $\hat{k'}$ by calculating the angle of refraction $\theta'$ from the angle of incidence $\theta$. This is given by
\begin{equation*}
    \theta'=\arcsin{\frac{n_{\rm in}}{n_{\rm out}} \sin{\theta}}\:\text{sign}{(\hat{N}\times \hat{k}\cdot \hat{z})}\:\text{sign}{(\hat{N}\cdot \hat{k})},
\end{equation*}
where $n_{\rm in}$ and $n_{\rm out}$ are the inside and outside refractive indices relative to the boundary curve, and the $\text{sign}$ terms are used to calculate the direction of the rotation correctly. Thus, we can finally predict the deflected direction $\hat{k'}$ using
\begin{equation*}
    \hat{k}' = \begin{cases}
    R(\theta')\hat{N}\text{sign}(\hat{k}\cdot \hat{N})&, \text{ if } \theta < \theta_{\rm c}\\
    (\hat{k}\cdot \hat{T}) \hat{T} - \hat{N}&, \text{ if } \theta \geq \theta_{\rm c}\\
    \end{cases},
\end{equation*}
where $\theta_{\rm c} = \arcsin \left( {n_{\rm out}/n_{\rm in}} \right)$ is the critical angle, and $R(\theta)$ is a counterclockwise rotation matrix. 

Now that we have a mathematical model for the surface of the lens that is adjustable using the set of control points $S$, we can write a program that shoots parallel rays perpendicular to the lens and optimises the placement of the control points in order to minimise the focal spot within a specific distance range. Specifically, the cost function to minimise is determined as
\begin{equation*}
    f(\vec{\theta}) = 1 - \max_{\vec{r}}(\text{rate}(\vec{r},\vec{\theta}))\,,
\end{equation*}
where $\theta$ is the vector that contains the coordinates of the control points in $S$, and $\text{rate}(\vec{r},\vec{\theta})$ is the fraction of rays that hit a photosensor at position $\vec{r}$ over the total number of rays sent after having passed through a lens with parameters $\vec{\theta}$. To perform the minimisation we use an algorithm called differential evolution, which is an iterative, evolutionary method to solve the constrained global minimisation problem. We also use hardware acceleration to explore a wider volume in parameter space.

The results showed that the optimal lens is an aspheric lens with a maximum detection rate of $0.93$. We identified the closest aspheric lens available from Thorlabs and proceeded with testing it with a GEANT4 simulation.

\subsection{Simulation}
After having identified a commercially available lens, we obtained its CAD model and built a GEANT4 simulation to calculate a better estimate for the photon incidence rate on the photodetector. To do so, we place a stack of 23 bilayers each made by SiO$_2$ and Si$_3$N$_4$ with the same chirped nominal thicknesses of our experimental apparatus. All boundaries allow specular lobe reflections. Each surface is decomposed into microfacets according to the UNIFIED ground model where a random angle $\alpha$ is added to the reflection angle of the ray that is sampled from a Gaussian distribution with $\sigma_{\alpha} = 0.1$\,rad.

Each photon possesses a random polarisation angle sampled from a unified distribution and is generated anywhere in between the stack, with an initial energy of $E_{\gamma} = 1.65\,{\rm eV}$ (where the measured boost factor spectrum peaks) and a perpendicular velocity component giving a random angle on the order of $10^{-3}$\,rad. 

\begin{figure}[t]
    \centering
    \includegraphics[width=0.45\textwidth, trim= 0 0 0 0 cm]{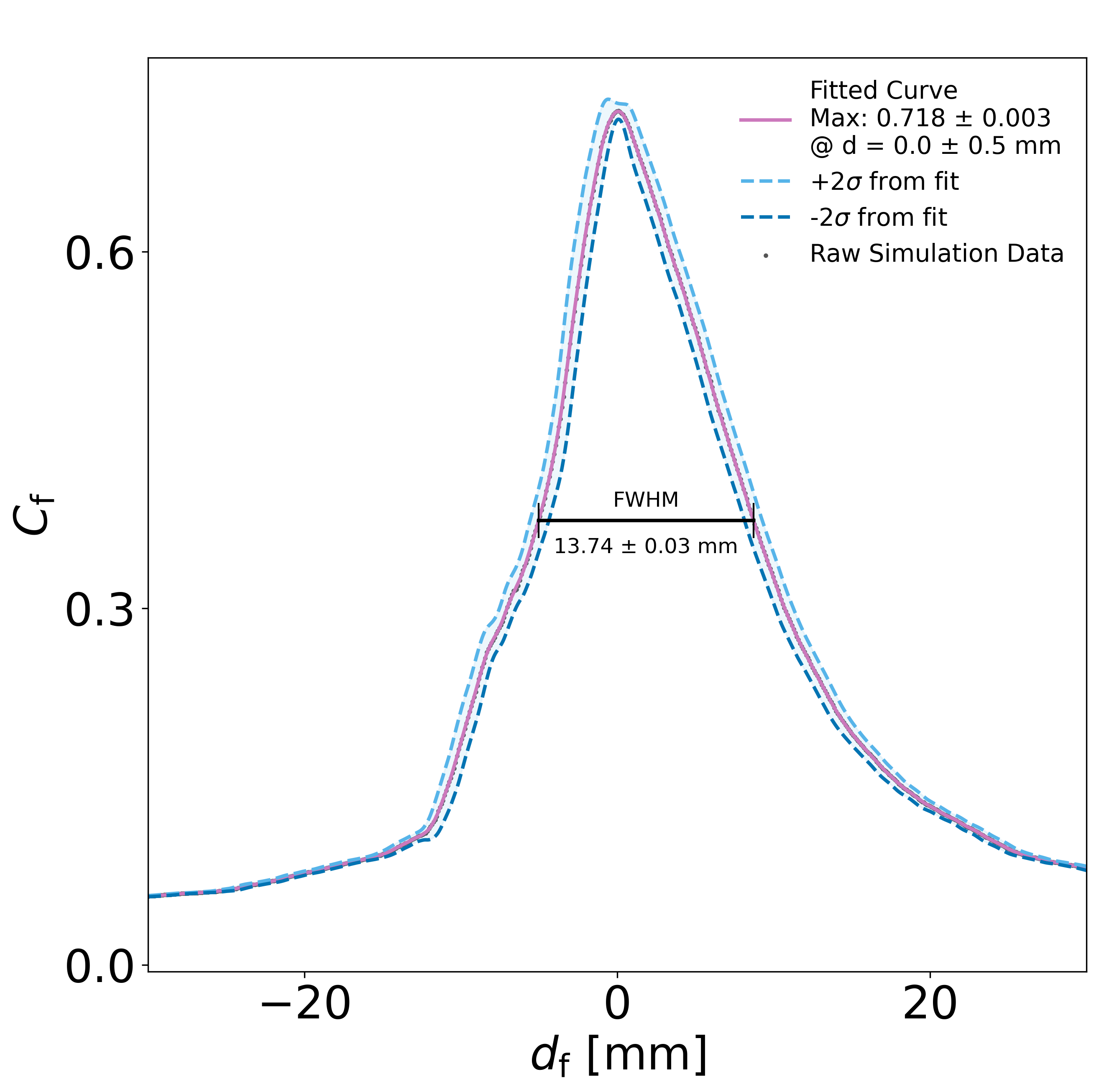}
    \caption{Simulated fraction of hits on photodetector ($C_{\rm f}$) as a function of displacement from the focal point of the selected physical lens ($d_{\rm f}$). The GEANT4 simulation takes into account other factors such as polarization, chromatic aberration, attenuation, and optical interface imperfections.}
    \label{fig:lens-GEANT4}
\end{figure}

Every material is assigned a refractive index as a function of the incident photon energy from which it is possible to calculate the reflectivity $R$ of each interface using
\begin{equation}
    R=\left|\frac{n_i-n_j}{n_i+n_j}\right|^2\,,
\end{equation}
where $n_i$ and $n_j$ refer to the refractive indices of two adjacent materials. 
With this model we are able to get the more accurate estimate of the photon incidence rate at the detector, which turns out to be 71.8\%. A plot of the fraction of hits ($C_{\rm f}$) at the focal point as a function of displacement from the focal point ($d_{\rm f}$) generated is shown in Fig.~\ref{fig:lens-GEANT4}.

\section{ \label{app:statistics} Detection of a signal in our counting experiment}
The experiment is carried out into two steps, with either the dielectric stack present (``on'') or removed (``off''). 
We call $\alpha$ the ratio between the period of time the ``on experiment'' is run and the ``off experiment'' is run, $t_{\rm on}/t_{\rm off}$, and the counts in each period \non and \noff.
The ``off experiment'' assumes that no dark photon signal is present without the stack which enables their conversion.

Such an experimental approach is the same as the one described in \cite{Vianello_2018}, i.e.~the detection of a source in a counting experiment. In this section we will define a test statistic and construct the upper limit based on the observed on- and off-counts.

\subsection{Upper limits for Poisson background}
\begin{figure}[t]
	\begin{center}
	\includegraphics[width=0.45\textwidth]{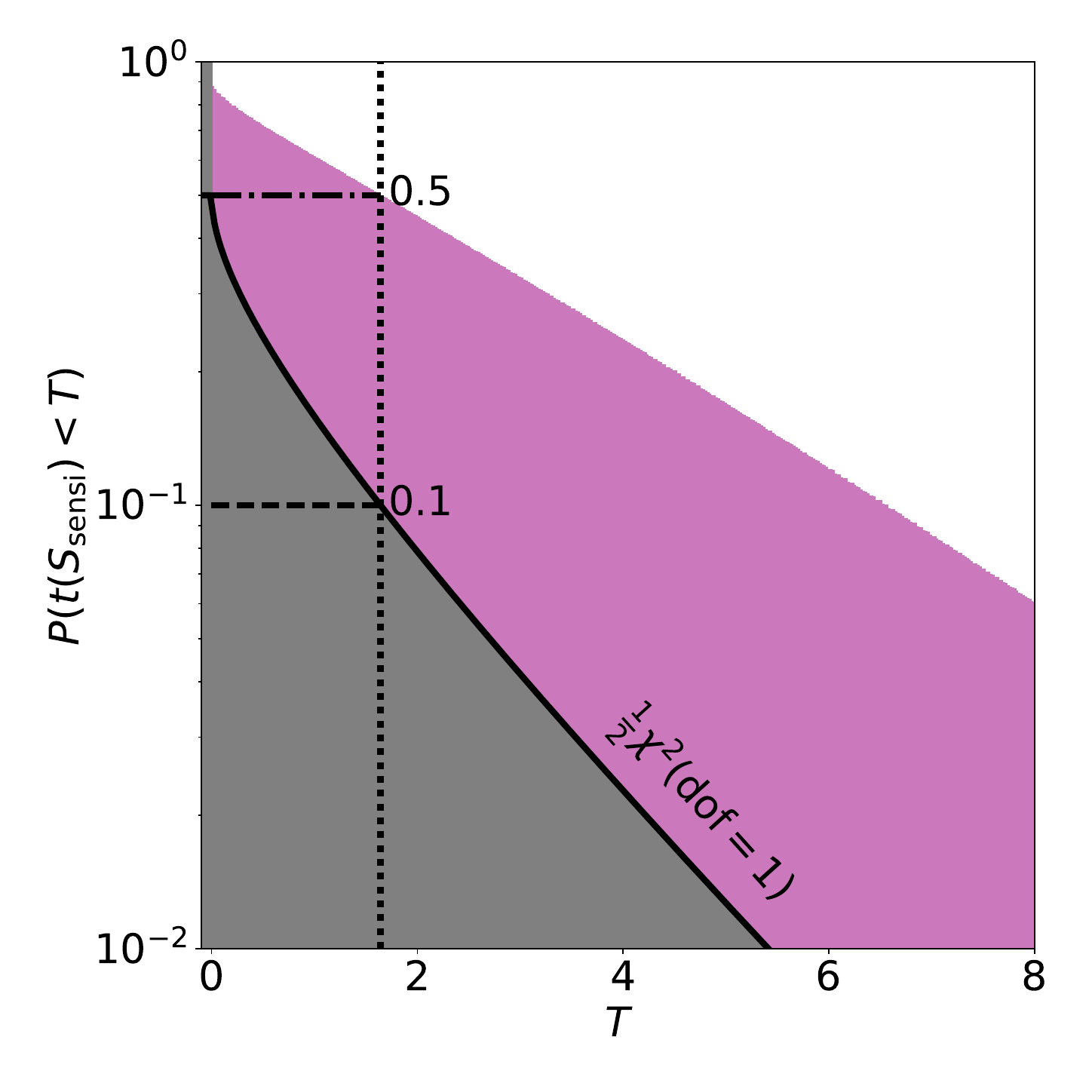} 
	\caption{Distribution of the test statistic $t(S_\mathrm{sensi})$ evaluated at $S=S_\mathrm{sensi}$, the median upper limit, given two hypotheses: either that $S=S_\mathrm{sensi}$ (gray histogram) or $S=0$ (magenta histogram). The figure shows that the median $t(S_\mathrm{sensi})$ under the hypothesis $S=0$ corresponds to the 90$^{\rm th}$ percentile of $t(S_\mathrm{sensi})$, and also that if $S=S_\mathrm{sensi}$, $t(S_\mathrm{sensi})$ is distributed as $\frac{1}{2}\chi^2(\mathrm{dof}=1)$ according to Wilks' theorem.
	}
	\label{fig:ts_distribution}
	\end{center}
\end{figure}

For an expected signal $S$ and background expectation $\alpha B$ in the ``on measurement'', the total likelihood is:
\begin{equation}
\lik(S,B | \non, \noff) = P(\non | S+ \alpha B) \times P(\noff | B)\,,
\end{equation}
where we use a Poisson counting distribution 
\begin{equation}
P(N|\mu) = \frac{\mu^N}{N!}e^{-\mu}\,.
\end{equation}
Both $S$ and $B$ are constrained in the fits to be non-negative. The profiled likelihood ratio $\lambda(S)$ expresses the compatibility between the best fit and the model where $S$ is the true signal strength:
\begin{align}
    \lambda(S) = \frac{\lik(S,\hbc | \non, \noff)}{\lik(\hat{S},\hat{B} | \non, \noff)}\,,
\end{align}
where $\hat{S}$ and $\hat{B}$ are the signal and background values that maximise the likelihood, while \hbc is the background expectation that conditionally maximises the likelihood given a fixed $S$.
The likelihood ratio test statistic we use for inference is given by:
\begin{equation}
    t(S) = -2 \ln \lambda(S)\,.
    \label{eq:ts}
\end{equation}
Expanding this expression and evaluating it for the hypothesis $S=S_0$ gives~\cite{LiMa_1983}

\begin{align}
\begin{split}
    t(S_0) &= 2 \bigg[ \non \ln{\frac{\alpha \hat{B}+\hat{S}}{\alpha \hbc+S_0}} + \noff \ln{\frac{\hat{B}}{ \hbc }} + \\ & - (\hat{S}-S_0) -(\alpha+1)(\hat{B}-B_c) \bigg]
\end{split}
\end{align}

We decided \emph{a priori} that we would report an upper-limit only result of this search. Therefore, we set $t(S) = 0$ when $S<\hat{S}$. Furthermore, both $S$ and $B$ are constrained to be non-negative in the fit. The best-fit signal and background rates can be computed analytically. 
The best-fit signal is
\begin{align}
    \hat{S} = \sup(N_\mathrm{on} - \alpha N_\mathrm{off},0)\,,
\end{align}
while the best-fit background given a signal is the positive root of
\begin{align}
a \cdot {\hat{B}(S)}^2 + b \cdot \hat{B}(S) + c = 0
\end{align}
where 
\begin{align}
\begin{split}
    a &= \alpha(\alpha +1) \\
    b &= (\alpha+1) S - \alpha (N_{\rm on} + N_{\rm off}) \\
    c &= -N_{\rm off} S\,.
\end{split}
\end{align}
Solving this for either $S=\hat{S}$ or $S=S_0$ gives the best-fit $\hat{B}$ or the conditional \hbc, respectively. 

Computing confidence intervals requires a comparison between $t_\mathrm{data}(S_0)$ computed for the experimental data for a certain signal $S_0$, and the cumulative distribution $F(t(S_0) | S=S_0)$, i.e.~the distribution function of $t(S_0)$ under the hypothesis that the true signal is $S_0$. Solving $1-F(t_\mathrm{data}(S) | S) = 0.1$ for S yields the $90\%$ confidence level upper limit on the signal strength. 
Figure~\ref{fig:ts_distribution} shows the distribution (shown as $1-F$) of $F(t(S_\mathrm{sensi}) | S=S_\mathrm{sensi})$ (magenta) and $F(t(S_\mathrm{sensi}) | S=0)$ (black), where $S_\mathrm{sensi}$ is the median upper limit on the signal. In the large-sample limit, $1-F(t_\mathrm{data}(S) | S)$ will tend towards a half-chi-squared distribution, as predicted by Wilks' theorem~\cite{wilks}. From the figure, it is clear that this regime is reached when the limit is around the experimental sensitivity for a $90\%$ confidence interval.


\bibliography{main}

\end{document}